\documentclass[preprint,12pt]{elsarticle}

\usepackage{graphicx}
\usepackage{amsmath,amssymb,amsfonts}
\usepackage{color}
\usepackage{algorithmic}
\usepackage{graphicx}
\usepackage{textcomp}
\usepackage{comment,color}
\usepackage{enumitem}
\usepackage{epsfig}
\usepackage{caption}
\usepackage{verbatim,bm,multicol,blindtext,subfigure}
\usepackage{epstopdf,array,mathtools,url}

\usepackage{lineno}

\journal{Elsevier}

\newcolumntype{L}{>{\centering\arraybackslash}m{3cm}}
\DeclarePairedDelimiter\norm{\lVert}{\rVert}%
\makeatletter
\let\oldabs\abs
\def\abs{\@ifstar{\oldabs}{\oldabs*}}

\let\oldnorm\norm
\def\norm{\@ifstar{\oldnorm}{\oldnorm*}}
\makeatother
\begin{document}

\begin{frontmatter}


\title{A holistic approach to computing first-arrival traveltimes using neural networks}



\author[1]{Umair bin Waheed}
\ead{umair.waheed@kfupm.edu.sa}
\address[1]{Department of Geosciences, King Fahd University of Petroleum and Minerals, Dhahran 31261,  Saudi Arabia.}


\author[2]{Tariq Alkhalifah}

\address[2]{Physical Sciences and Engineering Division, King
Abdullah University of Science and Technology, Thuwal 23955, Saudi Arabia.}

\author[3]{Ehsan Haghighat}
\address[3]{Department of Civil Engineering, Massachusetts Institute of Technology, 
MA 02139, USA.}

\author[2]{Chao Song}

\begin{abstract}
Since the original algorithm by John Vidale in 1988 to numerically solve the isotropic eikonal equation, there has been tremendous progress on the topic addressing an array of challenges, including improvement of the solution accuracy, incorporation of surface topography, adding more accurate physics by accounting for anisotropy/attenuation in the medium, and speeding up computations using multiple CPUs and GPUs. Despite these advances, there is no mechanism in these algorithms to carry information gained by solving one problem to the next. Moreover, these approaches may breakdown for certain complex forms of the eikonal equation, requiring {\color{black} simplification of the equations to estimate approximate solutions}. Therefore, we seek an alternate approach to address the challenge in a holistic manner, i.e., a method that not only makes it simpler to incorporate topography, allows accounting for any level of complexity in physics, benefiting from computational speedup due to the availability of multiple CPUs or GPUs, but also able to transfer knowledge gained from solving one problem to the next. We develop an algorithm based on the emerging paradigm of physics-informed neural network to solve various forms of the eikonal equation. We show how transfer learning and surrogate modeling can be used to speed up computations by utilizing information gained from prior solutions. We also propose a two-stage optimization scheme to expedite the training process in the presence of sharper heterogeneity in the velocity model and recommend using a locally adaptive activation function for faster convergence. Furthermore, we demonstrate how the proposed approach makes it simpler to incorporate additional physics and other features in contrast to conventional methods that took years and often decades to make these advances. Such an approach not only makes the implementation of eikonal solvers much simpler but also puts us on a much faster path to progress. The method paves the pathway to solving complex forms of the eikonal equation that have remained unsolved using conventional algorithms or solved using some approximation techniques at best; thereby, creating new possibilities for advancement in the field of numerical eikonal solvers. 
\end{abstract}


\end{frontmatter}


\section{Introduction}
\label{section1}

{\color{black} The eikonal equation is a nonlinear partial differential equation (PDE) obtained from the first term of the Wentzel-Kramers-Brillouin expansion of the wave equation and represents a class of Hamilton-Jacobi equations~\cite{crandall1983viscosity}. It finds applications in multiple domains of science and engineering, including image processing~\cite[]{alvino2007efficient}, robotic path planning and navigation~\cite[]{garrido2016path}, computer graphics~\cite[]{raviv2011affine}, and semi-conductor manufacturing~\cite[]{helmsen1996two}. In seismology, it is used to compute first-arrival traveltimes, which are necessary for the success of a wide range of seismic processing and imaging tools including statics and moveout correction~\cite{lawton1989computation}, traveltime tomography for initial velocity model building~\cite{hole19953,taillandier2009first}, microseismic source localization~\cite{grechka2015relative}, and ray-based migration~\cite{lambare20033d}.} Ray tracing and finite-difference based solutions of the eikonal equation are the most popular approaches for computing traveltimes. 

Ray tracing methods compute traveltimes along the characteristics of the eikonal equation by solving a system of ordinary differential equations ~\cite{cerveny2001seismic}. The approach is generally efficient for a sparse source-receiver geometry, but the computational cost increases dramatically with the increase in the number of source-receiver pairs. Moreover, for practical applications such as imaging and velocity model building, traveltime solutions need to be interpolated onto a regular grid. This requirement not only adds to the computational cost of the method but also poses a challenge, particularly in complex media where rays may diverge from one another, leading to large spatial gaps between rays, creating regions known as shadow zones~\cite{vidale1990finite}. Additionally, in strongly varying velocity models, multiple ray-paths may connect a source-receiver pair, making it easy to miss the path with the minimum traveltime. Therefore, the numerical solution of the eikonal equation has been a topic of continued research interest over the years.

\citeauthor{vidale1988finite}~\cite{vidale1988finite} led the development of numerical eikonal solvers by proposing an expanding box strategy to compute first-arrival traveltimes in heterogeneous media. Subsequently, the method was improved and extended to three dimensions~\cite{vidale1990finite}, to incorporate anisotropy~\cite{dellinger1991anisotropic,dellinger1997anisotropic}, and to high-order accurate solutions~\cite{kim19993}. The instability of the expanding box method due to turning rays led to the development of the expanding wavefront scheme~\cite{podvin1991finite}. This was further improved to obtain maximum energy traveltimes~\cite{nichols1996maximum}, and to incorporate anisotropy in the model~\cite{wang2006expanding}. 

Another algorithm that became popular during the late 1990s was the fast marching method~\cite{sethian19993}. The popularity of the method was due to its accuracy, stability, and efficiency properties. The fast marching method saw great interest and development in the subsequent period. This included extension of the method to improve traveltime accuracy~\cite{rickett1999second,alkhalifah2001implementing,popovici20023}, incorporating anisotropy~\cite{sethian2003ordered,cristiani2009fast,bin2015efficient}, parallelization for computational speedup using multiple CPUs~\cite{breuss2011adaptive}, and even acceleration using GPUs~\cite{monsegny2018fast}. 

Despite its success, the fast marching method was overtaken in popularity by the fast sweeping method~\cite{zhao2005fast} since the mid-2000s. This was mainly due to the flexibility and robustness of the fast sweeping method to various forms of the eikonal equation. Numerous advances to the fast sweeping method have since been proposed to improve the accuracy of the method~\cite{zhang2006high,fomel2009fast}, to incorporate anisotropy~\cite{luo2012fast,waheed2015iterative,han2017calculating,le2019discontinuous}, to account for attenuation~\cite{hao2018fast}, to tackle surface topography~\cite{lan2013high}, and parallelization for computational speedup~\cite{zhao2007parallel,detrixhe2013parallel}. 

Several other hybrid strategies have also been proposed to solve the eikonal equation. For a detailed review of these methods, we refer the interested reader to~\cite{gomez2019fast}.

In light of these developments, it is beyond doubt that there has been tremendous progress since the original eikonal solver by \citeauthor{vidale1988finite}~\cite{vidale1988finite}. This huge and growing body of literature, spanning over three decades, on the numerical solution of the eikonal equation, required significant research efforts to address an array of challenges, including improvement of the solution accuracy, incorporation of surface topography, adding more accurate physics by accounting for anisotropy/attenuation in the medium, and speeding up computations by using multiple CPUs and GPUs. Therefore, we seek an alternate approach that could address these challenges in a holistic manner -- a method that makes it simpler to incorporate topography, allow accounting for more accurate physics, and benefit from computational speedup due to the availability of multiple CPUs or GPUs. Such an approach would not only make the implementation of eikonal solvers much simpler but also put us on a much faster path to progress in solving complex forms of the eikonal equation.

Furthermore, a major drawback of the conventional eikonal solvers is that there is no mechanism to utilize the information gained by solving one problem to the next. Therefore, the same amount of computational effort is needed even for a small perturbation in the source position and/or the velocity model. This can lead to a computational bottleneck, particularly for imaging/inversion applications that require repeated computations, often with thousands of source positions and multiple updated velocity models. Therefore, a method that could use information gained from one solution to the next to speed up computations can potentially remedy this situation. With these objectives in mind, we look into the machine learning literature for inspiration. 

Having shown remarkable success across multiple research domains~\cite{jordan2015machine}, machine learning has recently shown promise in tackling problems in scientific computing. The idea to use an artificial neural network for solving PDEs has been around since the 1990s~\cite{lagaris1998artificial}. However, due to recent advances in the theory of deep learning coupled with a massive increase in computational power and efficient graph-based implementation of new algorithms and automatic differentiation, we are witnessing a resurgence of interest in using neural networks to approximate solutions of PDEs. 

{\color{black} Recently, Raissi et al.~\cite{raissi2019physics} developed a deep learning framework for the solution and discovery of PDEs. The so-called physics-informed neural network (PINN) leverages the capabilities of deep neural networks (DNNs) as universal function approximators. Contrary to the conventional deep learning approaches, PINNs restrict the space of admissible solutions by enforcing the validity of the underlying PDE governing the actual physics of the problem. This is achieved by using a simple feed-forward neural network leveraging automatic differentiation (AD) to compute the differential variables in the PDE. It is worth noting that PINNs do not require a \emph{labeled data} to learn the mapping between inputs and outputs, rather learning is facilitated through the loss function formed by the underlying PDE. PINNs have already demonstrated success in solving forward and inverse problems in geophysics~\cite[]{bin2020eikonal,smith2020eikonet,moseley2020solving,song2021solving,waheed2021pinntomo}. Unlike classical discretization-based methods, PINNs are the only unified framework that can be used readily both for data-driven and PDE-based forward and inverse solution of Eikonal equations by incorporating different terms in the loss function.}

In this chapter, we present a neural network approach to solve various forms of the eikonal equation. We use the PINN framework, where the governing equation is incorporated into the loss function of the neural network. We also show how the proposed method addresses the highlighted challenges compared to conventional algorithms. Specifically, we show that by simply updating the loss function of the neural network, we can account for more accurate physics in the traveltime solution. Moreover, since the proposed method is mesh-free, we will observe that to incorporate topography, no special treatment is needed as opposed to conventional finite-difference methods. In addition, the use of computational graphs allows us to run the same piece of code on different platforms (CPUs, GPUs) and architectures (desktops, clusters) without worrying about the implementation details. Most importantly, the proposed method allows us to use information gained while solving for a particular source position and velocity model to speed up computations for perturbations in the velocity model and/or source position. We demonstrate this aspect through the use of machine learning techniques like transfer learning and surrogate modeling.

The rest of the chapter is organized as follows: We begin by presenting the theoretical foundations of the proposed method and discuss how it can be used to solve more complex forms of the eikonal equation. Next, we test the method on a diverse set of 2D and 3D benchmark synthetic models and compare its performance with the popular fast sweeping method. Finally, we conclude the chapter by discussing the strengths of the method and identifying future research opportunities.

\section{Theory}

In this section, we describe how neural networks can be used to compute traveltime solutions for eikonal equations corresponding to isotropic and anisotropic media. We do so by first introducing the different forms of the eikonal equation and the concept of factorization. Next, we outline the general mechanism of a feed-forward neural network followed by its capability as a function approximator. This is followed by a brief overview of the concept of automatic differentiation, which is used to compute the derivative of the networks' output with respect to the inputs. Finally, putting these concepts together, we will present the proposed algorithm for solving various forms of the eikonal equation.

\subsection{Eikonal equations}
\label{section: 2.1}

The eikonal equation is a non-linear first-order PDE that is, for an isotropic medium, given as:
\begin{equation}
    \left(\frac{\partial T}{\partial x}\right)^2 + \left(\frac{\partial T}{\partial z}\right)^2 = \frac{1}{v(x,z)^2},
\label{eq:isoeikonal}
\end{equation}
subject to a point-source boundary condition as:
\begin{equation}
       T(x_s,z_s)  = 0,
\end{equation}
where $T(x,z)$ is the traveltime from the source point $(x_s,z_s)$ to a point $(x,z)$ in the computational domain, and $v(x,z)$ is the phase velocity of the isotropic medium. 
Since the curvature of the wavefront near the point-source is extremely large, previous studies~\cite{fomel2009fast,waheed2017fast} have shown that it is better to solve the factored eikonal equation instead of equation~\eqref{eq:isoeikonal}. The idea is to factor the unknown traveltime into two multiplicative factors, where one of the factors is specified analytically to capture the source-singularity such that the other factor is gently varying in the source neighborhood. Therefore, we factor $T(x,z)$ into two multiplicative functions:
\begin{equation}
    T(x,z) = T_0(x,z) \cdot \tau(x,z),
    \label{eq:factorization}
\end{equation}
where $T_0(x,z)$ is the known function and $\tau(x,z)$ is the unknown function. Plugging this into equation~\eqref{eq:isoeikonal}, we get the factored eikonal equation for an isotropic model as: 
\begin{equation}
    \left(T_0\frac{\partial \tau}{\partial x} + \tau\frac{\partial T_0}{\partial x}\right)^2 + \left(T_0\frac{\partial \tau}{\partial z} + \tau\frac{\partial T_0}{\partial z}\right)^2 = \frac{1}{v^2},
\label{eq:facisoeikonal}
\end{equation}
subject to the updated point-source condition:
\begin{equation}
       \tau(x_s,z_s)  = 1.
\label{eq:bc}
\end{equation}
The known factor $T_0$ is the traveltime solution in a homogeneous isotropic model given as:
\begin{equation}
    T_0(x,z) = \frac{\sqrt{(x-x_s)^2 + (z-z_s)^2}}{v_s},
    \label{eq:homisosol}
\end{equation}
where $v_s$ is taken to be the velocity at the point-source location.

\begin{figure}[]
\begin{center}
\includegraphics[width=0.98\textwidth]{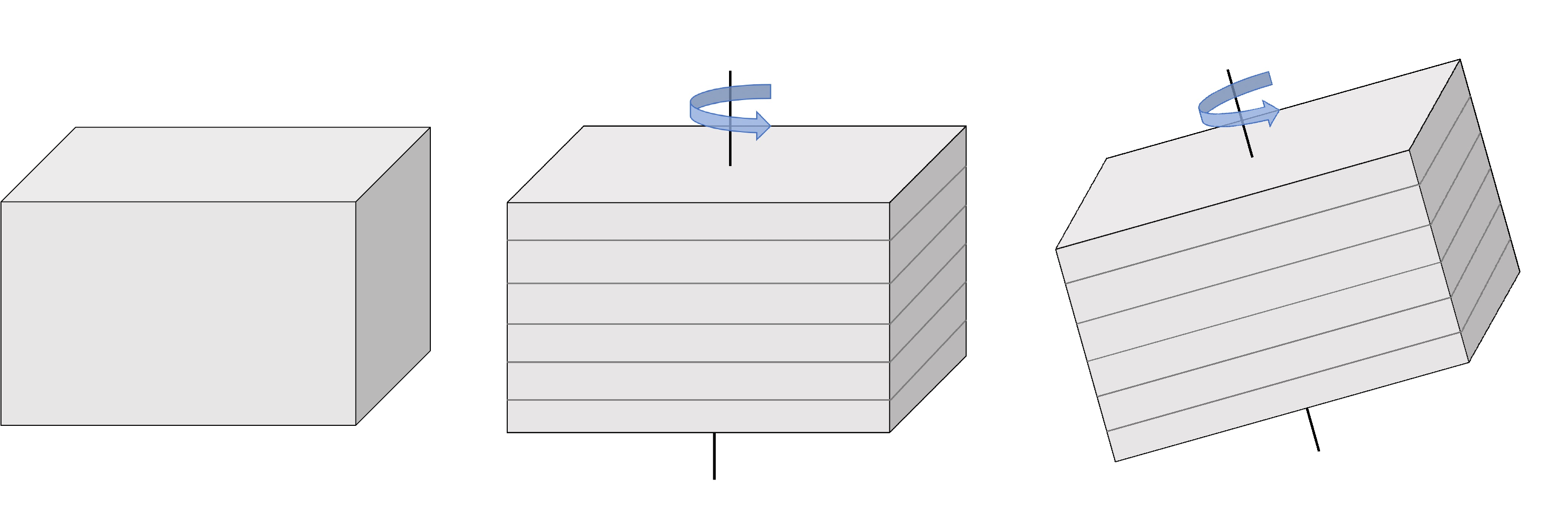}
\end{center}
\caption{
{\color{black} Illustration of the different subsurface approximations: isotropic (left), vertically transversely isotropic (VTI) (center), and tilted transversely isotropic (TTI) (right). The solid lines indicate the direction of the symmetry axis for the VTI and TTI cases.}
}%
\label{fig:anisotropies}
\end{figure}

Equation~\eqref{eq:facisoeikonal} is the factored eikonal equation for an isotropic medium; however, sedimentary rocks exhibit at least some degree of anisotropy due to a number of factors including thin layering and preferential alignment of grains cracks ~\cite{thomsen1986weak}. This results in the velocity being a function of the wave propagation direction, making the isotropic approximation of the Earth invalid. Therefore, traveltime computation algorithms must honor the anisotropic nature of the Earth for accurate subsurface imaging and other applications. Thus, we consider a realistic approximation of the subsurface anisotropy known as the tilted transverse isotropy (TTI) case. The factored eikonal equation for a TTI medium is considerably more complex than the isotropic case and is given, under the acoustic assumption, as~\cite{waheed2017fast}:
\begin{equation}
\resizebox{0.93\hsize}{!}{$
\begin{aligned}
     & (1+2\epsilon)\left(\cos\theta\left(T_0\frac{\partial \tau}{\partial x} + \tau\frac{\partial T_0}{\partial x}\right) + \sin\theta\left(T_0\frac{\partial \tau}{\partial z} + \tau\frac{\partial T_0}{\partial z}\right)\right)^2 \\
     + & \left(\cos\theta\left(T_0\frac{\partial \tau}{\partial z} + \tau\frac{\partial T_0}{\partial z}\right) - \sin\theta\left(T_0\frac{\partial \tau}{\partial x} + \tau\frac{\partial T_0}{\partial x}\right)\right)^2 \\
     \times & \left(1-\frac{2\eta v_t^2 (1+2\epsilon)}{1+2\eta}\left(\cos\theta\left(T_0\frac{\partial \tau}{\partial x} + \tau\frac{\partial T_0}{\partial x}\right) + \sin\theta\left(T_0\frac{\partial \tau}{\partial z} + \tau\frac{\partial T_0}{\partial z}\right)\right)^2\right) = \frac{1}{v_t^2},
\label{eq:facttieikonal}
\end{aligned}
$}
\end{equation}
where $v_t(x,z)$ is the velocity along the symmetry axis, $\epsilon(x,z)$ and $\eta(x,z)$ are the anisotropy parameters, and $\theta(x,z)$ is the tilt angle that the symmetry axis makes with the vertical. The point-source condition is the same as the one given in equation~\eqref{eq:bc}. Again $\tau(x,z)$ is the unknown function we solve equation~\eqref{eq:facttieikonal} for, whereas $T_0(x,z)$ is the known function which may be taken as the solution of a homogeneous, tilted elliptically isotropic medium, given as~\cite{luo2012fast}:
\begin{equation}
    T_0(x,z) = \sqrt{\frac{b_s(x-x_s)^2 + 2 c_s(x-x_s)(z-z_s) + a_s(z-z_s)^2}{a_s b_s - c_s^2}},
    \label{eq:homteasol}
\end{equation}
with
\begin{equation}
    \begin{aligned}
    a_s & = v_{ts}^2(1+2\epsilon_s)\cos\theta_{s}^2 + v_{ts}^2\sin\theta_{s}^2,\\
    b_s & = v_{ts}^2\cos\theta_{s}^2 + v_{ts}^2(1+2\epsilon_s)\sin\theta_{s}^2,\\
    c_s & = \left(v_{ts}^2-v_{ts}^2(1+2\epsilon_s)\right)\cos\theta_{s}\sin\theta_{s}.    
    \end{aligned}
\end{equation}
In the above expressions, $v_{ts}$ and $\epsilon_s$ are the velocity along the symmetry axis and the anisotropy parameter, respectively, at the point-source location. Similarly, $\theta_s$ is the tilt angle taken at the source-point.

{\color{black} It is worth highlighting that the isotropic and TTI cases represent mathematical approximations of the subsurface. An isotropic model considers the velocity to be invariant with respect to the direction of propagation, which is a crude representation of the Earth's crust. The simplest practical anisotropic symmetry system is axisymmetric anisotropy, commonly known as transverse isotropy (TI). A TI medium with a vertical axis of symmetry (VTI) is a good approximation for horizontally layering shale formation or thin-layering sediments. The factored eikonal equation for VTI media can be obtained by setting $\theta = 0$ in equation~\eqref{eq:facttieikonal}. For more complex geology,  such as sediments near the flanks of salt domes and fold-and-thrust belts like the Canadian foothills, a TTI model represents the best approximation. Figure~\ref{fig:anisotropies} illustrates these approximations graphically.}

The reason for considering eikonal equations corresponding to different media is to highlight, in comparison with the conventional methods, how easy it is to adapt the proposed method to solve a relatively more complex eikonal equation (more on this in Section~\ref{section: 2.4}). 

\subsection{Approximation property of neural networks}
\label{section: 2.2}

A feed-forward neural network, also known as a multi-layer perceptron, is a set of neurons organized in layers in which evaluations are performed sequentially through the layers. It can be seen as a computational graph with an input layer, an output layer, and an arbitrary number of hidden layers. In a fully connected neural network, neurons in adjacent layers are connected with each other, but neurons within a single layer share no connections. It is called a feed-forward neural network because information flows from the input through each successive layer to the output. Moreover, there are no feedback or recursive loops in a feed-forward neural network. 

\begin{figure}[]
\begin{center}
\includegraphics[width=0.65\textwidth]{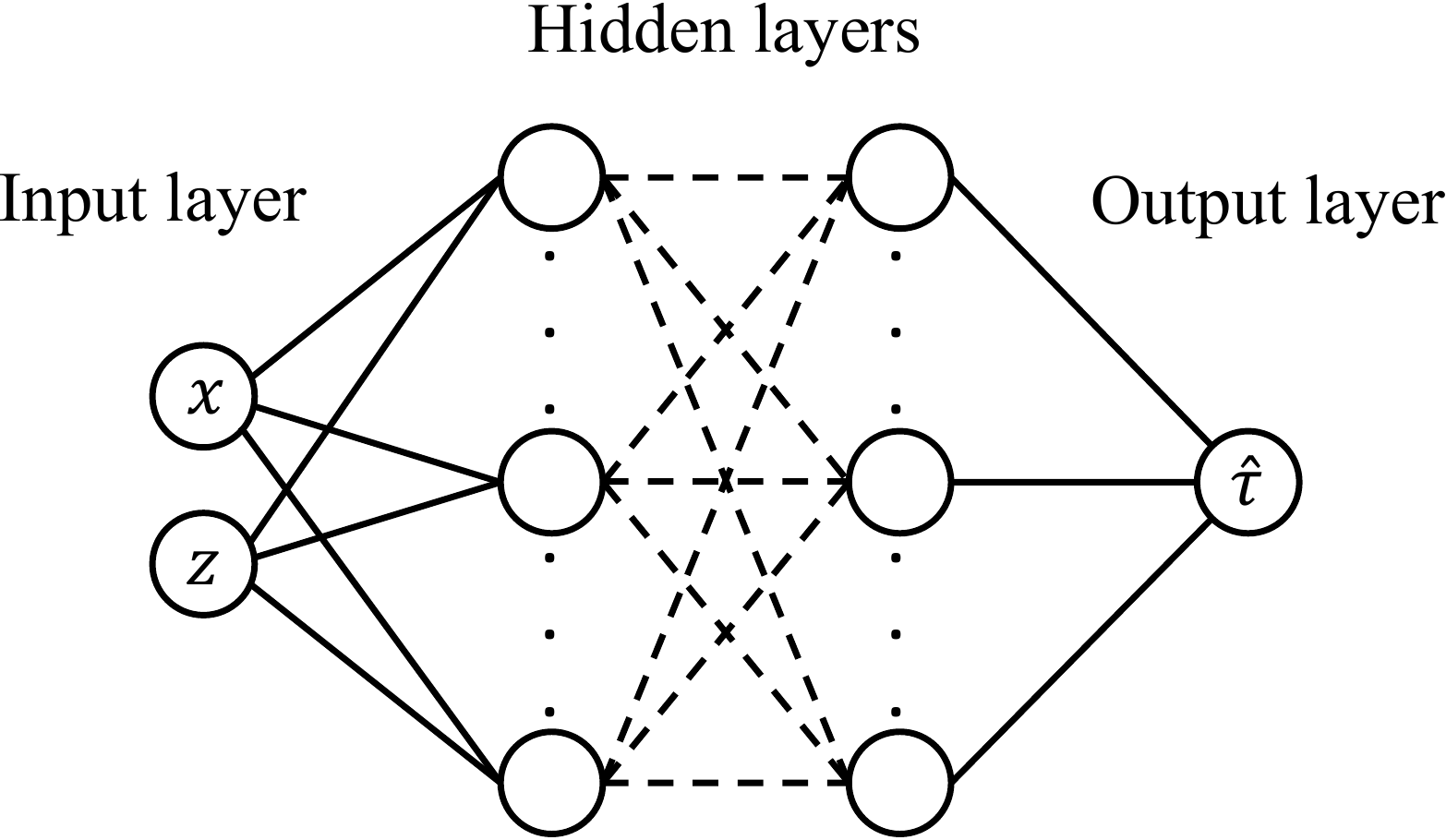}
\end{center}
\caption{
{\color{black} A feedforward neural network architecture containing an arbitrary number of hidden layers/neurons that is used to approximate the traveltime factor $\hat{\tau}$ at a given spatial coordinates $(x,z)$ for a 2D computational domain. }
}%
\label{fig:nnarch}
\end{figure}

Neural networks are well-known for their strong representational power. A neural network with $n$ neurons in the input layer and $m$ neurons in the output layer can be used to represent a function $u: \mathbb{R}^n \rightarrow \mathbb{R}^m$. In fact, it has been shown that a neural network with a finite number of neurons in the hidden layer can be used to represent any bounded continuous function to the desired accuracy. This is also known as the universal approximation theorem~\cite{cybenko1989approximation,hornik1989multilayer}. In addition, it was later shown that by using a deep network with multiple hidden layers and a nonlinear activation function, the total number of neurons needed to represent a given function could be significantly reduced~\cite{lu2017expressive}. Therefore, our goal here is to train a DNN that could represent the mapping between the spatial coordinates $(x,z)$, as inputs to the network, and the unknown traveltime function $\tau(x,z)$ representing the output of the DNN. {\color{black} Figure~\ref{fig:nnarch} illustrates this idea pictorially showing a neural network with input neurons for the spatial coordinates $(x,z)$ that are passed through the hidden layers to the output layer for predicting the traveltime factor at the inputted spatial location.} 

We formulate here considering a 2D case for simplicity of illustration. In a 3D model, one would need a neural network with three input neurons, one for each spatial dimension. It must also be noted that while DNNs are, in theory, capable of representing very complex functions, finding the actual parameters (weights and biases) needed to solve a given PDE can be very challenging.

\subsection{Automatic differentiation}
\label{section: 2.3}

Solving a PDE using neural networks requires a mechanism to accurately compute derivatives of the network's output(s) with respect to the input(s). There are multiple ways to compute derivatives including hand-coded analytical derivatives, symbolic differentiation, numerical approximation, and automatic differentiation (AD)~\cite{baydin2017automatic}. While manually working out the derivatives is exact, it is often time consuming to code and it is error-prone. Symbolic differentiation, while also exact, may result in exponentially large expressions and, therefore, can be prohibitively slow and memory intensive. Numerical differentiation, on the other hand, is easy to implement but can be highly inaccurate due to round-off errors. Contrary to these approaches, AD uses exact expressions with floating-point values instead of symbolic strings and it involves no approximation errors. This results in an accurate evaluation of derivatives at machine precision. Therefore, we evaluate the partial derivatives of the unknown traveltime factor $\tau$ with respect to the inputs $(x,z)$ using AD.

However, it must be noted that an efficient implementation of the AD algorithm can be non-trivial. Fortunately, many existing computational frameworks such as $\texttt{Tensorflow}$~\cite{tensorflow2015-whitepaper} and $\texttt{PyTorch}$~\cite{adam2017automatic} have made available efficiently implemented AD libraries. 

\subsection{Solving eikonal equations}
\label{section: 2.4}

We begin by considering how the different pieces of the puzzle outlined in the previous subsections can be combined to solve eikonal equations. First, we illustrate this using the factored isotropic eikonal equation~\eqref{eq:facisoeikonal} and then demonstrate how simple it is under the proposed framework to solve a more complex eikonal equation, such as the factored TTI eikonal equation~\eqref{eq:facttieikonal}.

To solve equation~\eqref{eq:facisoeikonal}, we leverage the capabilities of neural networks as function approximators and define a loss function that minimizes the residual of the underlying PDE for a chosen set of training (collocation) points. This is achieved using the following components:
\begin{itemize}
    \item [i.] a DNN approximation of the unknown traveltime field variable $\tau(x,z)$,
    \item [ii.] a differentiation algorithm, i.e., AD in this case, to evaluate partial derivatives of $\tau(x,z)$ with respect to the spatial coordinates $(x,z)$,
    \item [iii.] a loss function incorporating the underlying eikonal equation, sampled on a collocation grid, and
    \item [iv.] an optimizer to minimize the loss function by updating the neural network parameters.
\end{itemize}

To illustrate the idea, let us consider a two-dimensional domain $\Omega \in \mathbb{R}^2$. 
A point-source is located at coordinates $(x_s,z_s)$, where $\tau(x_s,z_s) = 1$. The unknown traveltime factor $\tau(x,z)$ is approximated using a DNN, $\mathcal{N}_\tau$, such that:

\begin{equation}
    \tau(x,z) \approx \hat{\tau}(x,z) = \mathcal{N}_\tau(x,z; \boldsymbol{\theta}),
\end{equation}
where $x,z$ are network inputs, $\hat{\tau}$ is the network output, and  $\boldsymbol{\theta}\in\mathbb{R}^D$ represents the set of all trainable parameters of the network with $D$ as the total number of parameters. 

The loss function is given by the mean-squared error norm as
\begin{equation}
\mathfrak{J} = \frac{1}{N_I}\sum_{(x^{*},z^{*}) \in I} \norm{\mathcal{L} }^2 + \frac{1}{N_I}\sum_{(x^{*},z^{*}) \in I} \norm{\mathcal{H}(-\hat{\tau})|\hat{\tau}|}^2 \\  
+ \norm{\hat{\tau}(x_s,z_s) - 1}^2,
\label{eq:loss_mse}
\end{equation}

where $\mathcal{L}$ represents the residual of the factored isotropic eikonal equation~\eqref{eq:facisoeikonal}, given by

\begin{equation}
\begin{aligned}
    \mathcal{L} = \left(T_0\frac{\partial \hat{\tau}}{\partial x} + \hat{\tau}\frac{\partial T_0}{\partial x}\right)^2 + \left(T_0\frac{\partial \hat{\tau}}{\partial z} + \hat{\tau}\frac{\partial T_0}{\partial z}\right)^2 - \frac{1}{v^2}.
\label{eq:iso_residual}
\end{aligned}
\end{equation}

The first term on the right side of equation~\eqref{eq:loss_mse} imposes validity of the factored eikonal equation~\eqref{eq:facisoeikonal} on a given set of training points $(x^{*},z^{*})~\in~I$, where $N_I$ is the number of training samples. The second term forces the solution $\hat{\tau}$ to be positive by penalizing negative solutions using the Heaviside function $\mathcal{H}()$. The last term enforces the boundary condition by imposing the solution $\hat{\tau}$ to be unity at the source point $(x_s,z_s)$. The set of network parameters $\boldsymbol{\theta}^*$ that minimizes the loss function \eqref{eq:loss_mse} on this set of training points, $(x^{*},z^{*})~\in~I$, is then identified by solving the optimization problem:
\begin{equation}
\boldsymbol{\theta}^* = \arg\min_{\boldsymbol{\theta}\in\mathbb{R}^D} \mathfrak{J}(x^*,z^*; \boldsymbol{\theta}).
\label{eq:optimization}
\end{equation}

Once the DNN is trained, we evaluate the network on a set of regular grid-points in the computational domain to obtain the unknown traveltime field. The final traveltime solution is obtained by multiplying it with the known traveltime part, i.e.,

\begin{equation}
    \hat{T}(x,z) = T_0(x,z) \cdot \hat{\tau}(x,z).
\end{equation}

\begin{figure}
\begin{center}
\includegraphics[width=0.99\textwidth]{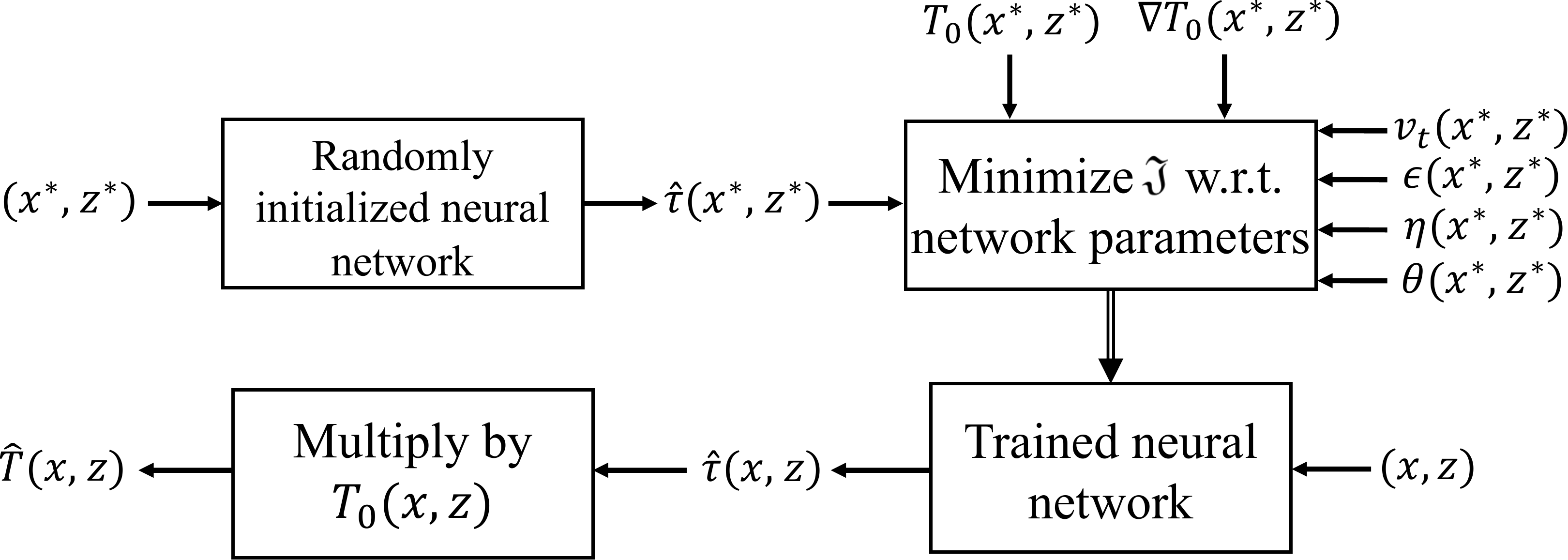}
\end{center}
\caption{
A workflow of the proposed factored TTI eikonal solver in 2D: A randomly initialized neural network is trained on a set of randomly selected collocation points $(x^{*},z^{*})$ in the model space with given model parameters $v_t(x^{*},z^{*})$, $\epsilon(x^{*},z^{*})$, $\eta(x^{*},z^{*})$, $\theta(x^{*},z^{*})$, the known traveltime function $T_0(x^{*},z^{*})$, and its spatial derivative $\nabla T_0(x^{*},z^{*})$ to minimize the loss function given in equation~\ref{eq:loss_mse}. Once the network is trained, it is evaluated on a regular grid of points $(x,z)$ to yield an estimate of the traveltime field $\hat{\tau}$, which is then multiplied with the factored traveltime part $T_0$ to yield the estimated first-arrival traveltime solution $\hat{T}$.
}%
\label{fig:flowchart}
\end{figure}

This yields traveltimes corresponding to an isotropic approximation of the Earth. However, it is well-known that the subsurface is anisotropic in nature. Therefore, a significant amount of research effort has been spent over the years on extending numerical eikonal solvers to anisotropic media. The complication in numerically solving the eikonal equation arises due to anellipticity of the wavefront~\cite{alkhalifah2000acoustic} resulting in high-order nonlinear terms in the eikonal equation. These high-order terms dramatically increase the complexity in solving the anisotropic eikonal equation and, therefore, have been a topic of immense research interest. On the contrary, the proposed neural network formulation allows solving for the anisotropic eikonal equation by simply replacing the residual in equation~\eqref{eq:loss_mse} with the one corresponding to the anisotropic eikonal equation. Therefore, to solve the factored TTI eikonal equation~\eqref{eq:facttieikonal}, we would, instead of equation~\eqref{eq:iso_residual}, use the following:

\begin{equation}
\resizebox{0.91\hsize}{!}{$
\begin{aligned}
     \mathcal{L} = & \, (1+2\epsilon)\left(\cos\theta\left(T_0\frac{\partial \hat{\tau}}{\partial x} + \hat{\tau}\frac{\partial T_0}{\partial x}\right) + \sin\theta\left(T_0\frac{\partial \hat{\tau}}{\partial z} + \hat{\tau}\frac{\partial T_0}{\partial z}\right)\right)^2 \\
     + & \,v_t^2\left(\cos\theta\left(T_0\frac{\partial \hat{\tau}}{\partial z} + \hat{\tau}\frac{\partial T_0}{\partial z}\right) - \sin\theta\left(T_0\frac{\partial \hat{\tau}}{\partial x} + \hat{\tau}\frac{\partial T_0}{\partial x}\right)\right)^2 \\
     \times & \left(1-\frac{2\eta v_t^2(1+2\epsilon)}{1+2\eta}\left(\cos\theta\left(T_0\frac{\partial \hat{\tau}}{\partial x} + \hat{\tau}\frac{\partial T_0}{\partial x}\right) + \sin\theta\left(T_0\frac{\partial \hat{\tau}}{\partial z} + \hat{\tau}\frac{\partial T_0}{\partial z}\right)\right)^2\right) - \frac{1}{v_t^2}.
\label{eq:tti_residual}
\end{aligned}
$}
\end{equation}
This is a highly desirable feature of this approach because eikonal equations corresponding to models with even lower symmetry than TTI can be easily solved by simply using a different residual term. By contrast, conventional algorithms such as fast marching or fast sweeping methods would require significant effort to incorporate such changes, thereby resulting in much slower scientific progress.

A workflow summarizing the proposed solver for the factored TTI eikonal equation is shown in Figure~\ref{fig:flowchart}.

\section{Numerical Tests}

In this section, we test the neural network based eikonal solver and compare its performance with the first-order fast sweeping method, which is routinely used in geophysical (and other) applications for traveltime computations. We consider several isotropic and anisotropic 2D/3D models for these tests and also include a model with topography to demonstrate the flexibility of the proposed method. 

For each of the examples presented below, we use a neural network having 20 hidden layers with 20 neurons in each layer and minimize the neural network's loss function using full-batch optimization. A locally adaptive inverse tangent activation function is used for all hidden layers except the final layer, which uses a linear activation function. Locally adaptive activation functions have been shown to achieve superior optimization performance and convergence speed over base methods. The introduction of a scalable parameter in the activation function for each neuron changes the slope of the activation function and, therefore, alters the loss landscape of the neural network for improved performance. For more information on locally adaptive activation functions, we refer the interested reader to~\cite{jagtap2020locally}. 

We choose the afore-mentioned configuration of the neural network based on some initial tests and keep it fixed for the entire study to minimize the need for hyper-parameter tuning for each new velocity model.

The following examples are prepared and trained using the neural network library, SciANN~\cite{haghighat2021sciann}, a Keras/Tensorflow API that is designed and optimized for physics-informed deep learning. SciANN leverages the latest advancements of Tensorflow while keeping the interface close to the mathematical description of the problem. 

\begin{figure}[]
\begin{center}
\includegraphics[width=0.55\textwidth]{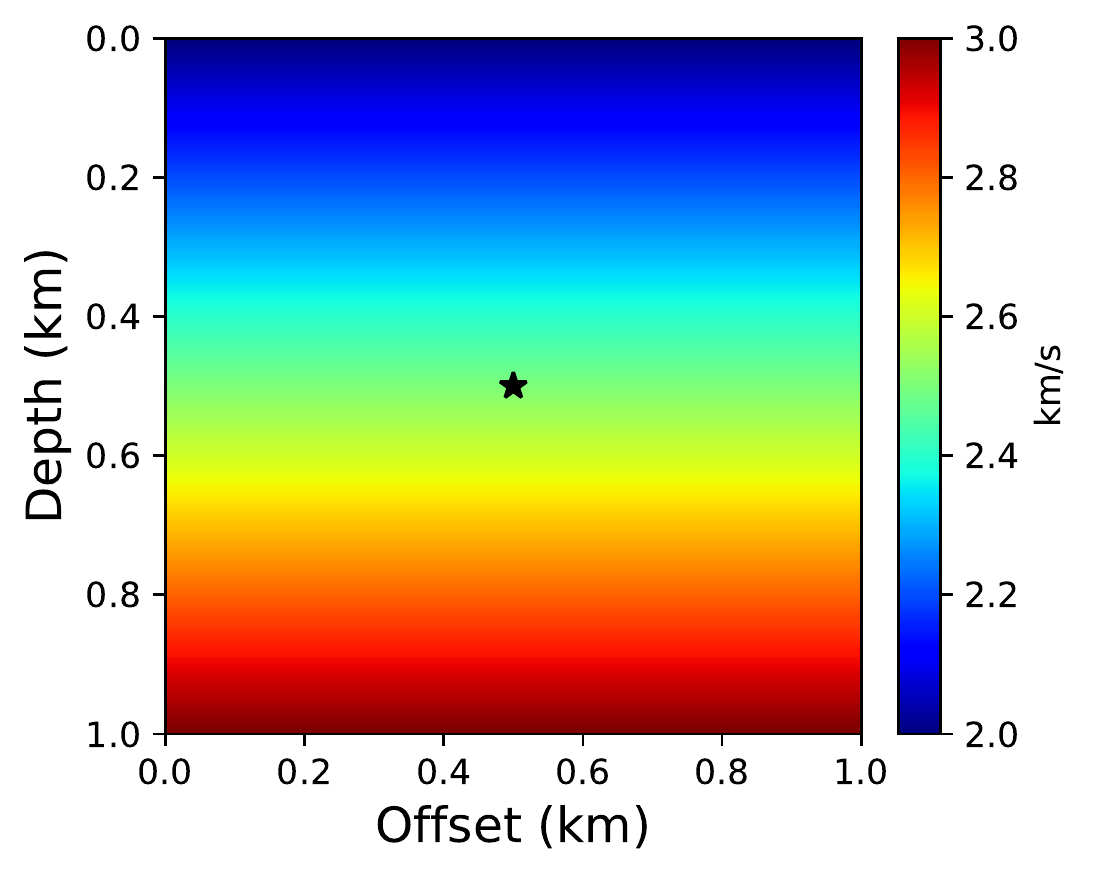}
\end{center}
\caption{
A vertically varying velocity model with a constant velocity gradient of 1~$\text{s}^{-1}$. The velocity at zero depth is equal to 2~km/s and it increases linearly to 3~km/s at a depth of 1~km. The black star indicates the point-source location used for the test.
}%
\label{fig:vofz_vz}
\end{figure}

\subsection*{Example 1: An isotropic model with constant vertically varying gradient}
First, we consider a vertically varying $1 \times 1$ km$^2$ isotropic model. The velocity at zero depth is 2~km/s and it increases linearly with a gradient of 1 s$^{-1}$ to 3~km/s at a depth of 1~km. We compute traveltime solutions using the neural network and first-order fast sweeping method by considering a point-source located at $(0.5~\text{km},0.5~\text{km})$. We compare their performance with a reference solution computed analytically~\cite{slotnick1959lessons}. The velocity model is shown in Figure~\ref{fig:vofz_vz} and is discretized on a 101$\times$101 grid with a 10 m grid interval along both axes.

\begin{figure}[]
\begin{center}
\subfigure[]{%
\label{fig:vofz_pinnerror}
\includegraphics[width=0.44\textwidth]{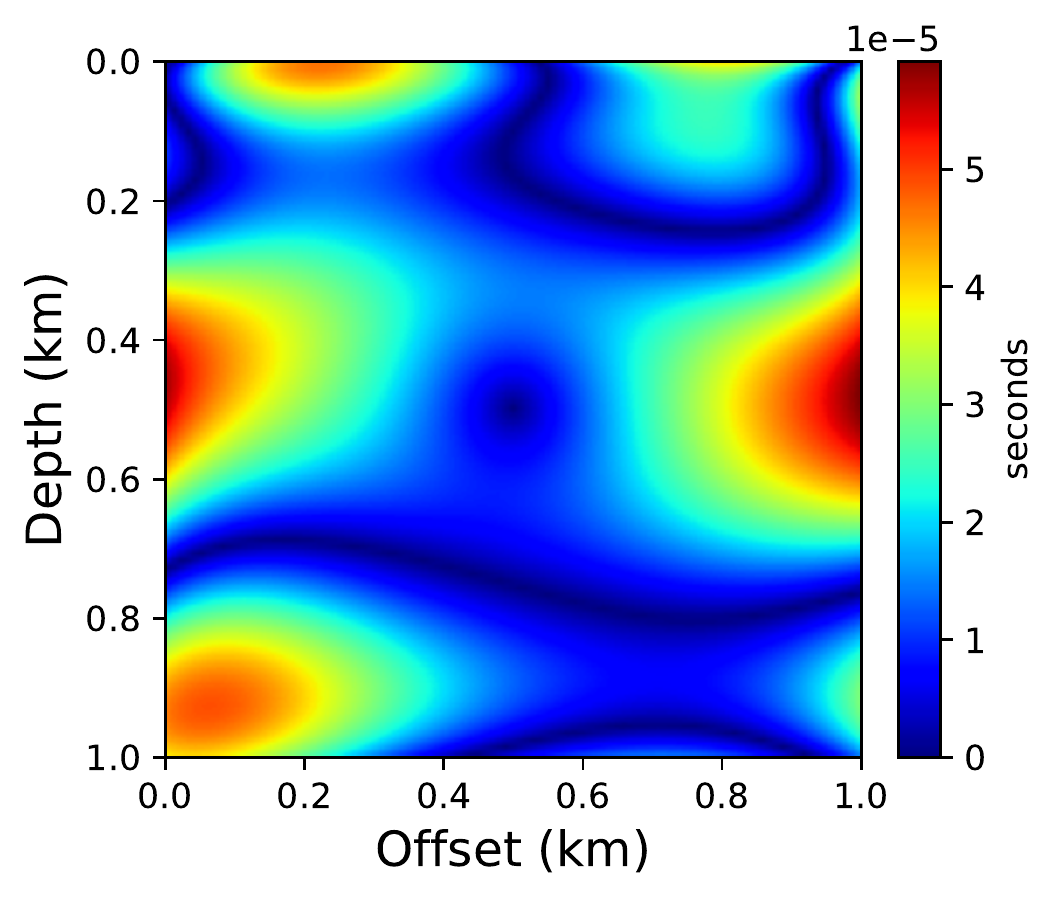}
}
\subfigure[]{%
\label{fig:vofz_fserror}
\includegraphics[width=0.47\textwidth]{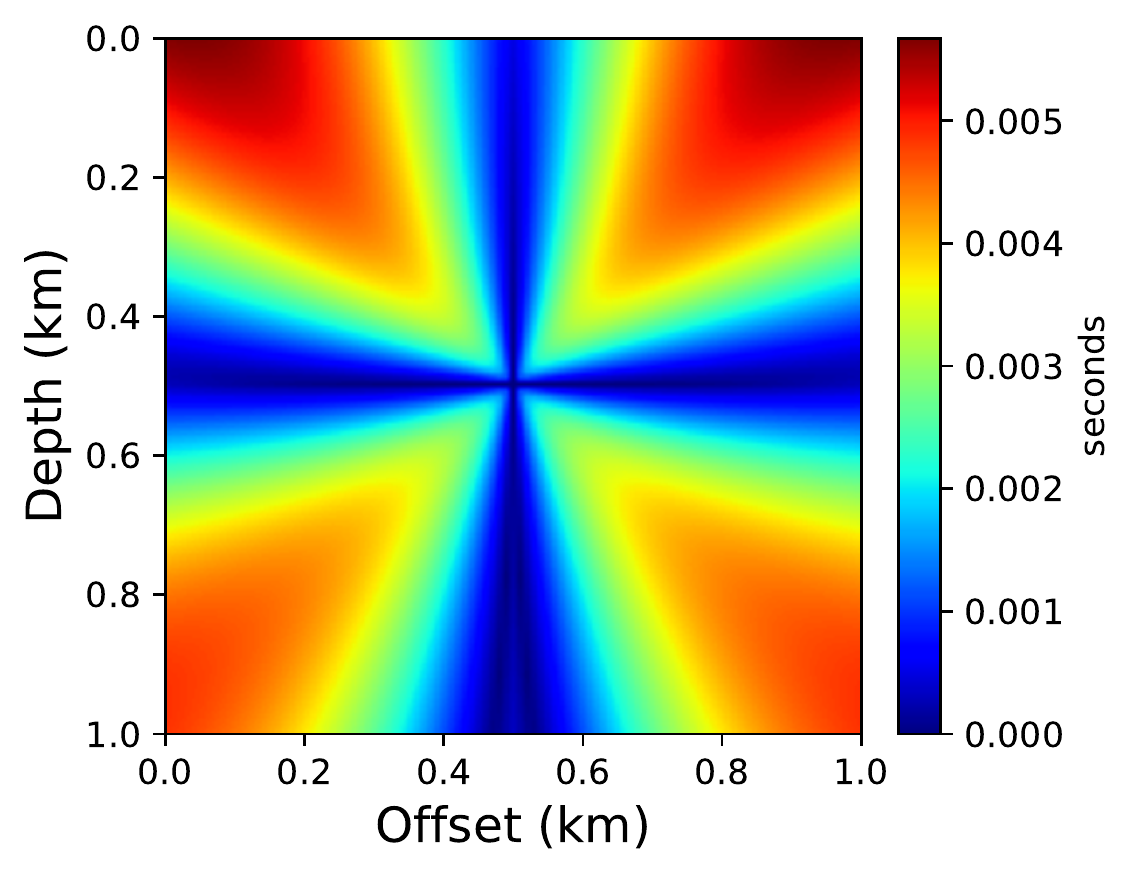}
}
\end{center}
\caption{
Absolute traveltime errors for the neural network solution (a) and the first-order fast sweeping solution (b) for the isotropic model and the source location shown in Figure~\ref{fig:vofz_vz}. 
}%
\label{fig:vofz_errors}
\end{figure}

\begin{figure}[]
\begin{center}
\includegraphics[width=0.5\textwidth]{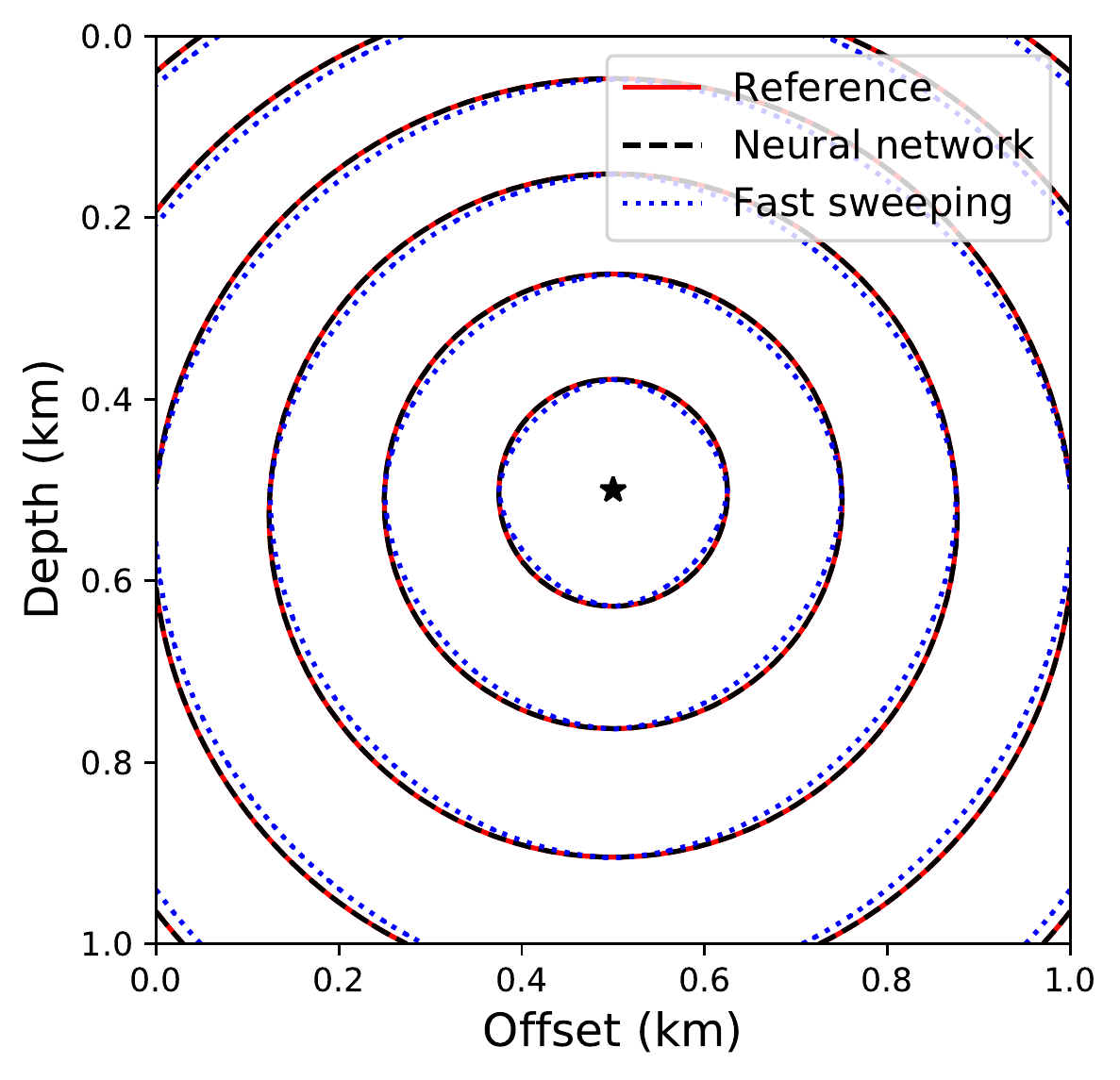}
\end{center}
\caption{
Traveltime contours for the reference solution (solid red) computed analytically, neural network solution (dashed black), and the first-order fast sweeping solution (dotted blue) for the vertically varying isotropic model. The black star shows the location of the point-source.
}%
\label{fig:vofz_contours}
\end{figure}

For training the neural network, we begin with randomly initialized parameters and train on 50\% of the total grid points selected randomly and use the Adam optimizer~\cite{kingma2014adam} with 10,000 epochs. Once the network is trained, we evaluate the trained network on the regularly sampled (101$\times$101) grid to obtain the unknown traveltime field $\hat{\tau}$, which is then multiplied with the corresponding factored traveltime field $T_0$ to obtain the final traveltime solution. We compare the accuracy of the neural network solution and the first-order fast sweeping solution, computed on the same regular grid in Figure~\ref{fig:vofz_errors}. We observe significantly better accuracy for the neural network based solution despite using only half of the total grid points for training. In Figure~\ref{fig:vofz_contours}, we confirm this observation by plotting the corresponding traveltime contours. 

\subsection*{Example 2: Smoothly varying TTI model}

\begin{figure}[]
\begin{center}
\includegraphics[width=0.55\textwidth]{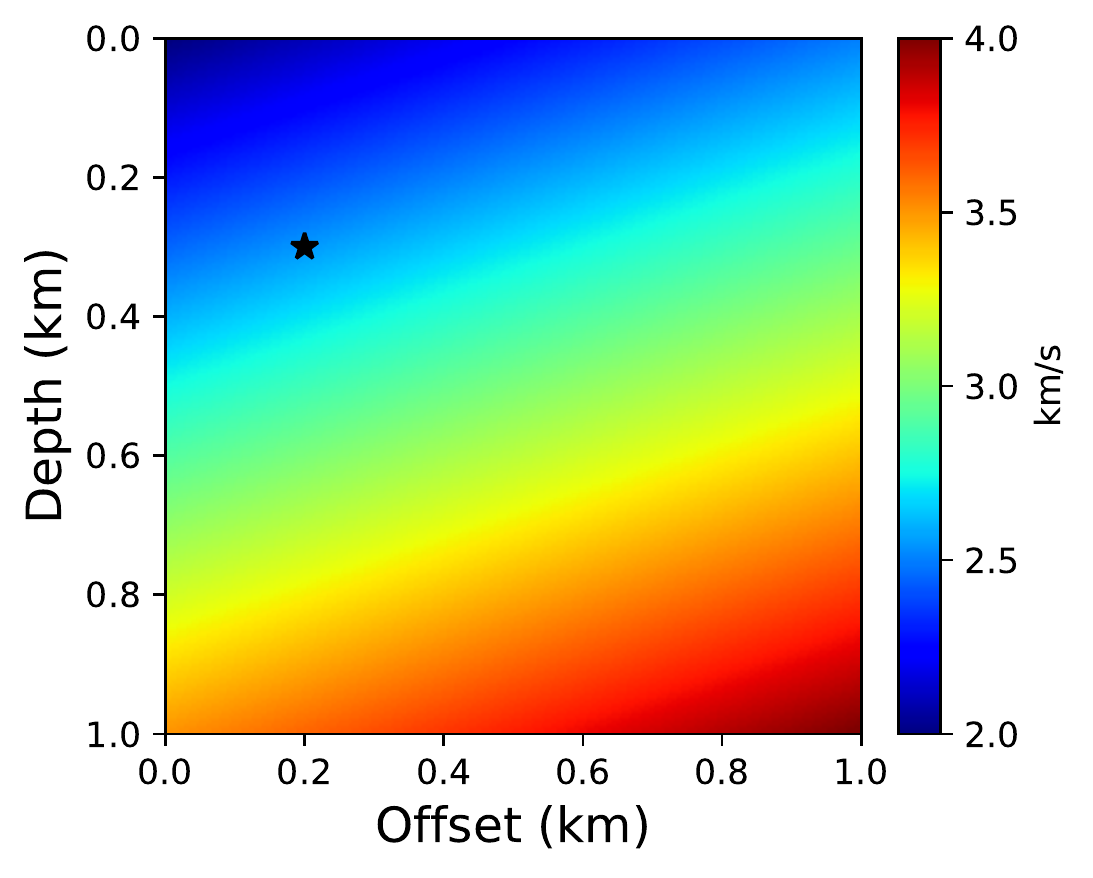}
\end{center}
\caption{
A velocity model for the parameter $v_t$ with a constant vertical gradient of 1.5~$\text{s}^{-1}$ and a horizontal velocity gradient of 0.5~$\text{s}^{-1}$. A homogeneous model is used for the anisotropy parameters ($\epsilon$=0.2, $\eta$ = 0.083) and for the tilt angle ($\theta$ = 30$^\circ$). The black star indicates the point-source location used for the test. 
}%
\label{fig:vofxz_vz}
\end{figure}

\begin{figure}[]
\begin{center}
\includegraphics[width=0.6\textwidth]{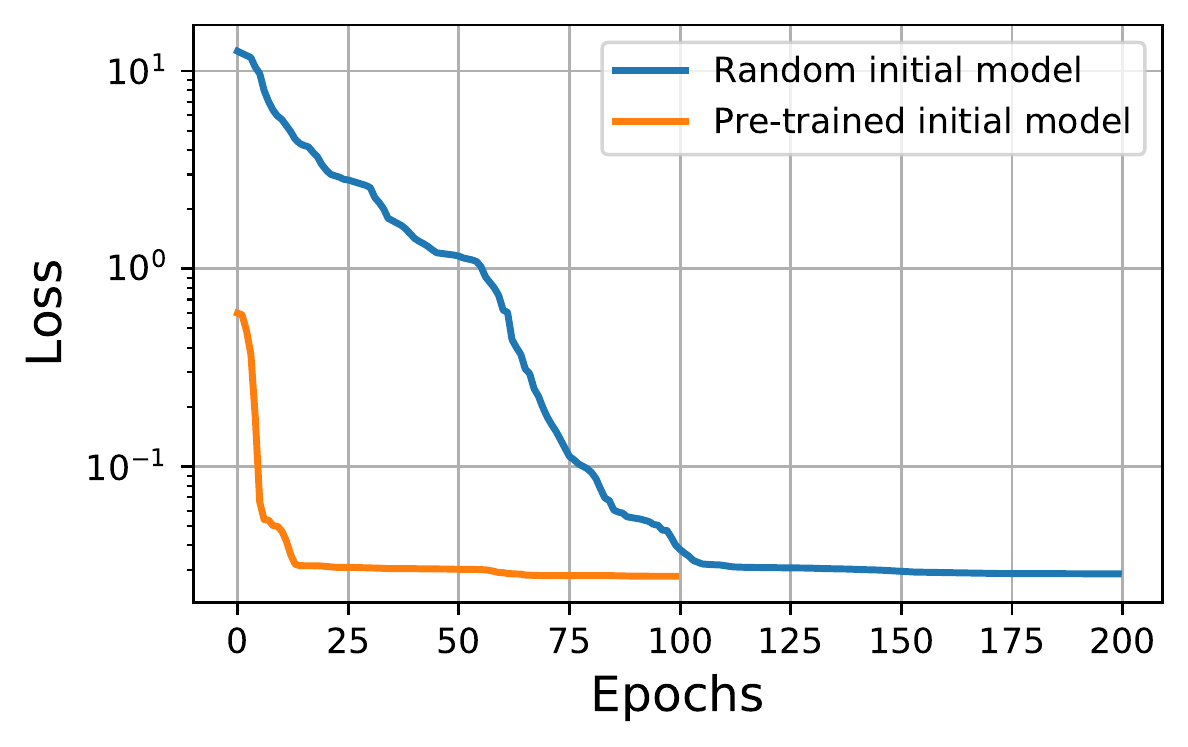}
\end{center}
\caption{
A comparison of the loss history for training of the TTI model in example 2 using pre-trained weights from example 1 (orange) and random initialization (blue). An L-BFGS-B optimizer is used for both cases.
}%
\label{fig:vofxz_transfer_loss}
\end{figure}

Next, we train the neural network to solve the TTI eikonal equation. Compared to the fast sweeping method that requires significant modifications to the isotropic eikonal solver, the neural network based approach requires only an update to the loss function by incorporating the appropriate residual based on the TTI eikonal equation. For the velocity parameter $v_t$, we consider a linear velocity model with a vertical gradient of 1.5~$\text{s}^{-1}$ and a horizontal gradient of 0.5~$\text{s}^{-1}$ as shown in Figure~\ref{fig:vofxz_vz}. We use homogeneous models for the anisotropy parameters with $\epsilon$=0.2 and $\eta$ = 0.083. We also consider a homogeneous tilt angle of $\theta$ = 30$^\circ$. These models are also discretized on a 101$\times$101 grid with 10 m grid interval along both axes.

\begin{figure}[]
\begin{center}
\subfigure[]{%
\label{fig:vofxz_pinnerror}
\includegraphics[width=0.47\textwidth]{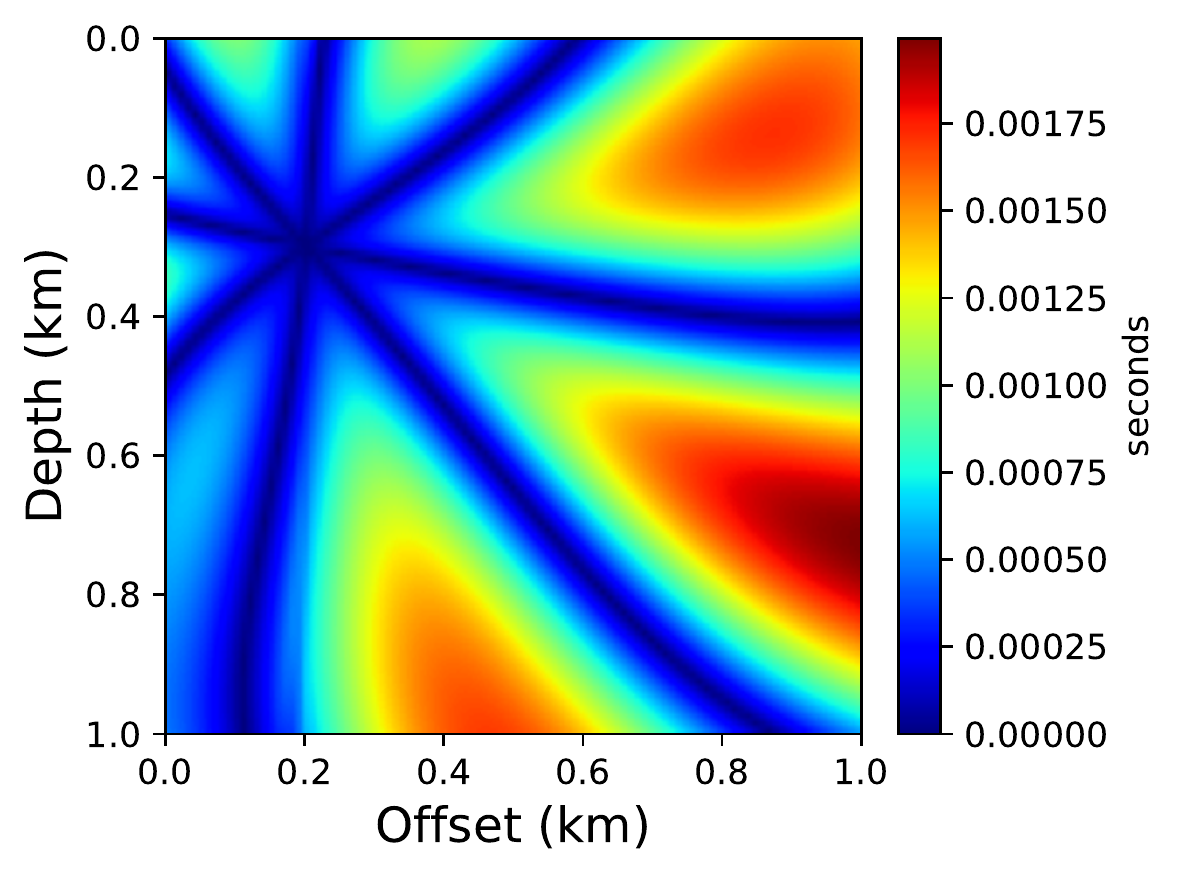}
}
\subfigure[]{%
\label{fig:vofxz_fserror}
\includegraphics[width=0.445\textwidth]{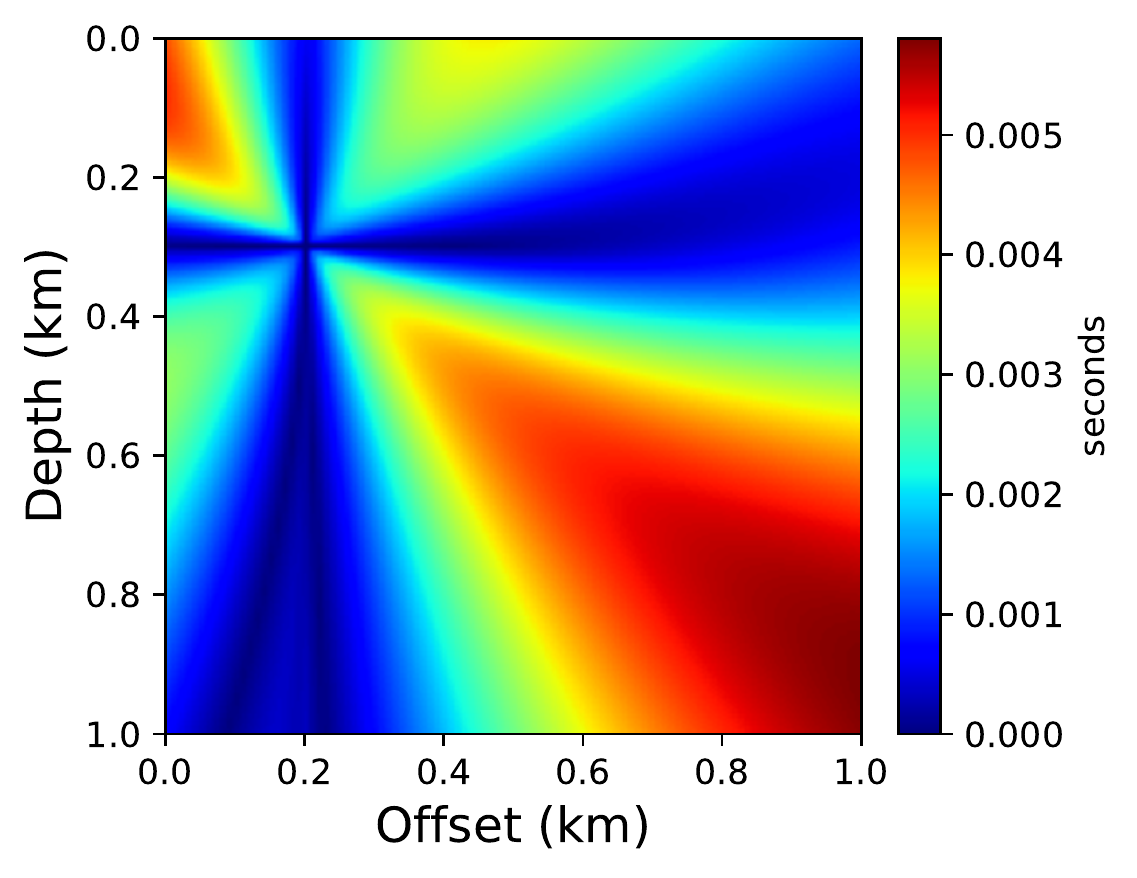}
}
\end{center}
\caption{
The absolute traveltime errors for the neural network solution (a) and the fast sweeping solution (b) for the TTI model considered in example 2.
}%
\label{fig:vofxz_errors}
\end{figure}

\begin{figure}[]
\begin{center}
\includegraphics[width=0.5\textwidth]{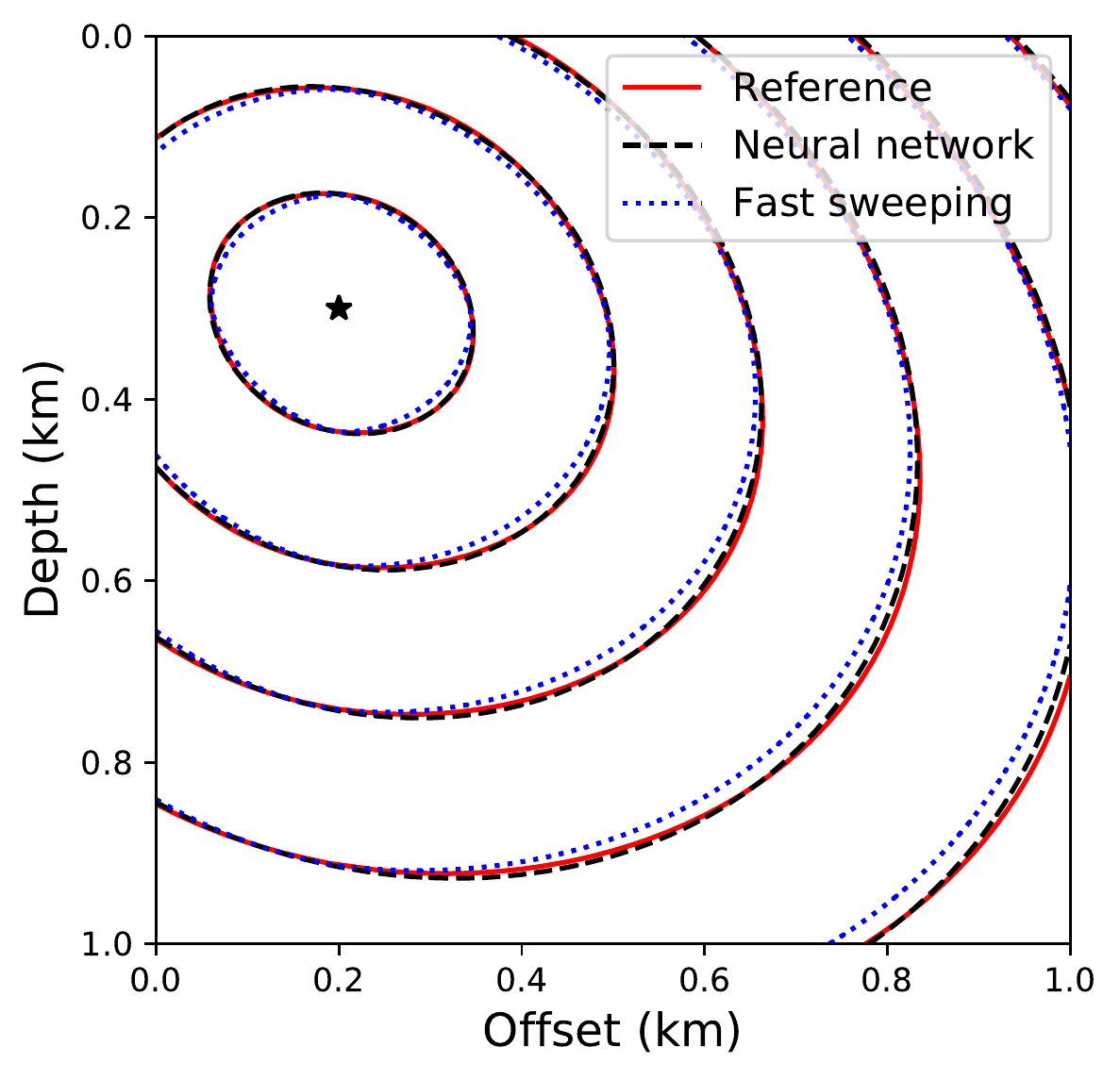}
\end{center}
\caption{
The traveltime contours for the reference solution (solid red), neural network solution (dashed black), and the first-order fast sweeping solution (dotted blue) for the smoothly varying TTI model. The black star shows the location of the point-source.
}%
\label{fig:vofxz_contours}
\end{figure}

Instead of training the network from scratch, we use transfer learning, which is a machine learning technique that relies on storing knowledge gained while solving one problem and applying it to a different but related problem. Starting with a pre-trained network from example 1, we fine-tune the neural network parameters for the TTI model using 50\% of the total grid points, selected randomly, using the L-BFGS-B solver~\cite{zhu1997algorithm} for 100 epochs. Starting with a pre-trained network allows us to use the L-BFGS-B method for faster convergence as opposed to starting with the Adam optimizer and then switching to L-BFGS-B as suggested by previous studies~\cite{raissi2019physics}. For comparison, we also train a neural network from scratch and the convergence history, shown in Figure~\ref{fig:vofxz_transfer_loss}, confirms that the solution converges much faster when using transfer learning.

Figure~\ref{fig:vofxz_errors} compares absolute traveltime errors computed using the neural network and the first-order fast sweeping method using the iterative solver of ~\citeauthor{waheed2015iterative}~\cite{waheed2015iterative}. The reference solution is obtained using a high-order fast sweeping method on a finer grid. We observe that, despite using transfer learning, the accuracy of the neural network solution is considerably better than the fast sweeping method. We confirm this observation visually by comparing the corresponding traveltime contours in Figure~\ref{fig:vofxz_contours}. {\color{black} One can also observe the effect of the additional anisotropy parameters and the tilt angle on the traveltime contours here. By comparing the shapes of the contours in Figure~\ref{fig:vofz_contours} and~\ref{fig:vofxz_contours}, it is obvious that the wave propagation speed varies with the direction of propagation. A faster propagation is observed orthogonal to the symmetry direction, given by the tilt angle, compared to the propagation along the symmetry axis.}

It is worth noting that while the complexity of the fast sweeping solvers and their computational cost increases dramatically when switching from an isotropic to a TTI model, for the neural network both cases require similar complexity and computational cost. Therefore, the proposed method is particularly suited to model complex physics involving media with anisotropy, attenuation, etc.

One of the main challenges in seismic imaging and inversion is the need for repeated traveltime computations for thousands of source locations and multiple updated velocity models. Unfortunately, conventional techniques do not allow the transfer of information from one solution to the next and, therefore, the same amount of computational effort is needed for even small perturbations in the source location and/or the velocity model. We noted above how transfer learning can be used to speed up convergence for a new velocity model and source position. This could be further extended by adding source location $(x_s,z_s)$ as input to the network and training a surrogate model.

\begin{figure}[]
\begin{center}
\includegraphics[width=0.55\textwidth]{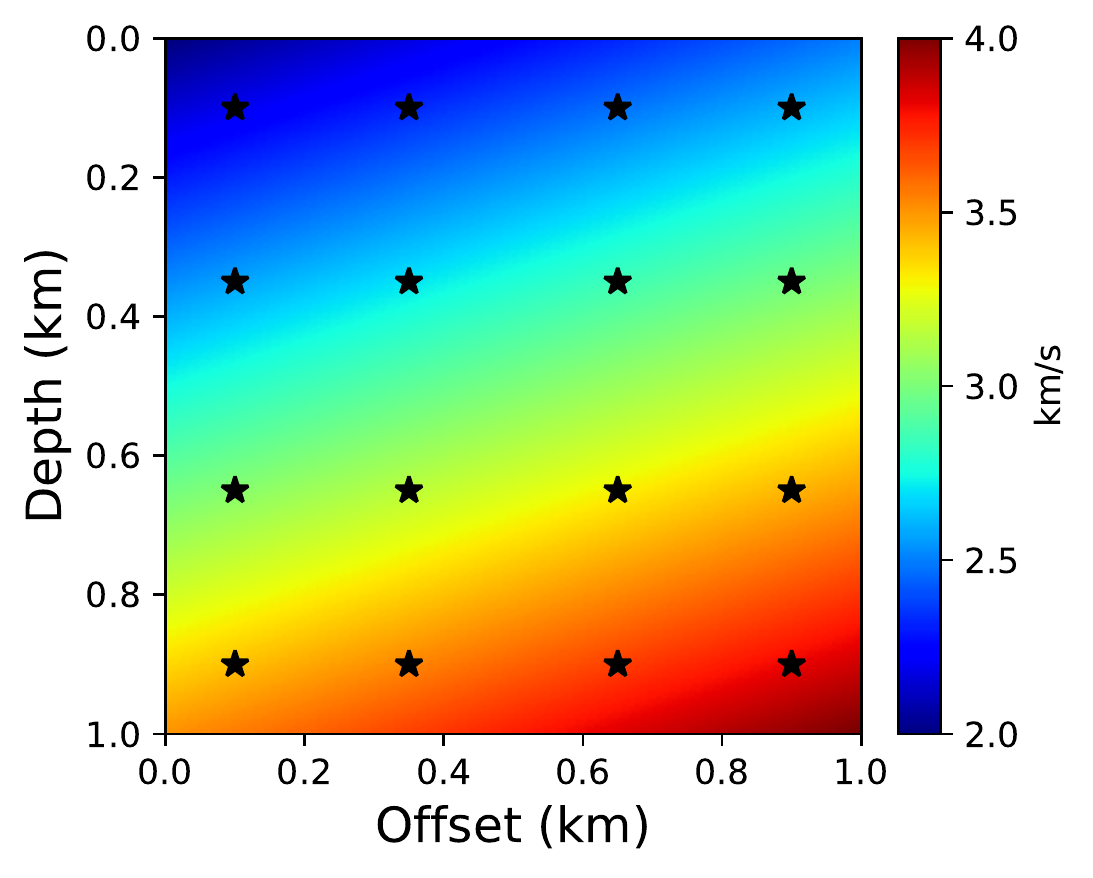}
\end{center}
\caption{
A velocity model for the parameter $v_t$ with a constant vertical gradient of 1.5~$\text{s}^{-1}$ and a horizontal velocity gradient of 0.5~$\text{s}^{-1}$. A homogeneous model is used for the anisotropy parameters ($\epsilon$=0.2, $\eta$ = 0.083) and for the tilt angle ($\theta$ = 30$^\circ$). Black stars indicate locations of sources used to train the network as a surrogate model.
}%
\label{fig:vofxz_vz_sur}
\end{figure}

To do so, we train a neural network on solutions computed for 16 sources located at regular intervals in the considered TTI model as shown in Figure~\ref{fig:vofxz_vz_sur}. Through the training process, the network learns the mapping between a given source location and the corresponding traveltime field. Once the surrogate model is trained, traveltime fields for additional source locations can be computed instantly by using a single evaluation of the trained network. This is similar to obtaining an analytic solver as no further training is needed for computing traveltimes corresponding to new source locations. This feature is particularly advantageous for large 3D models that need thousands of such computations.

\begin{figure}[]
\begin{center}
\subfigure[]{%
\label{fig:vofxz_sur_pinnerror}
\includegraphics[width=0.46\textwidth]{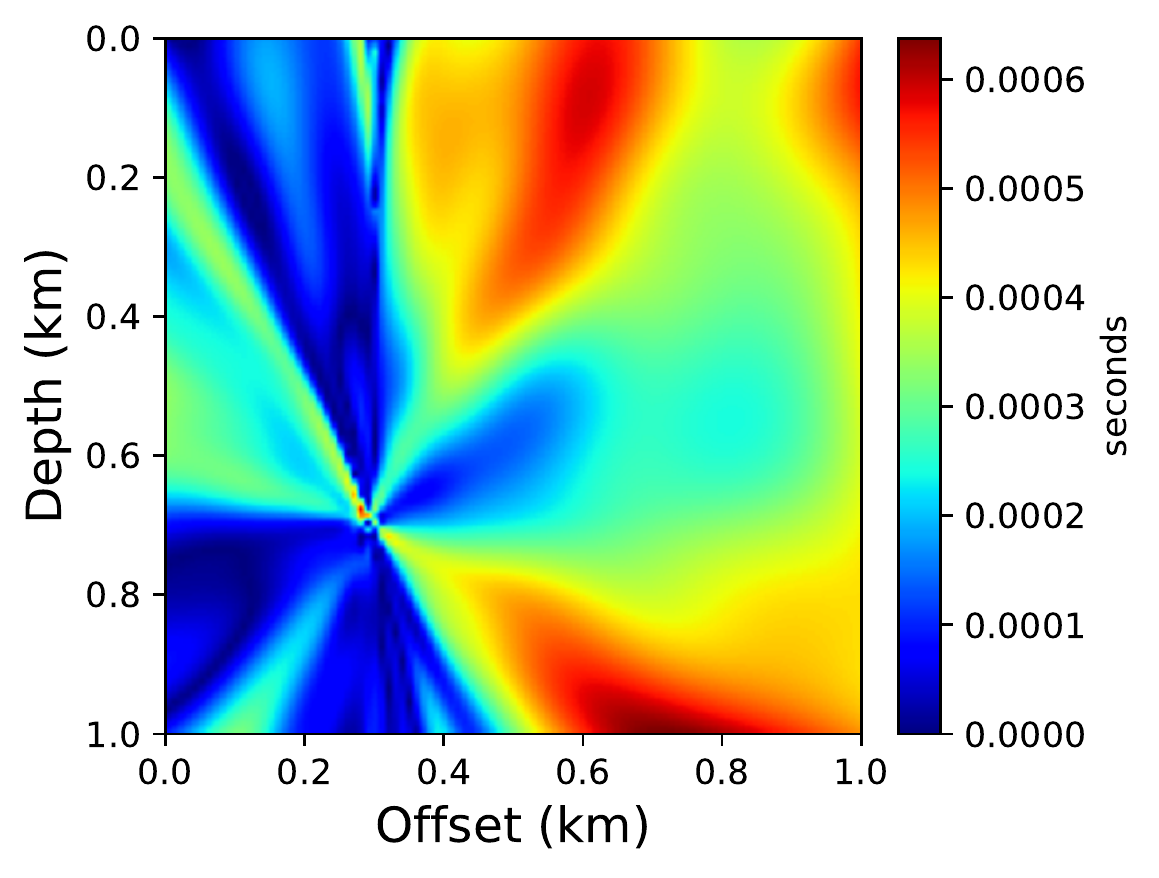}
}
\subfigure[]{%
\label{fig:vofz_sur_fserror}
\includegraphics[width=0.46\textwidth]{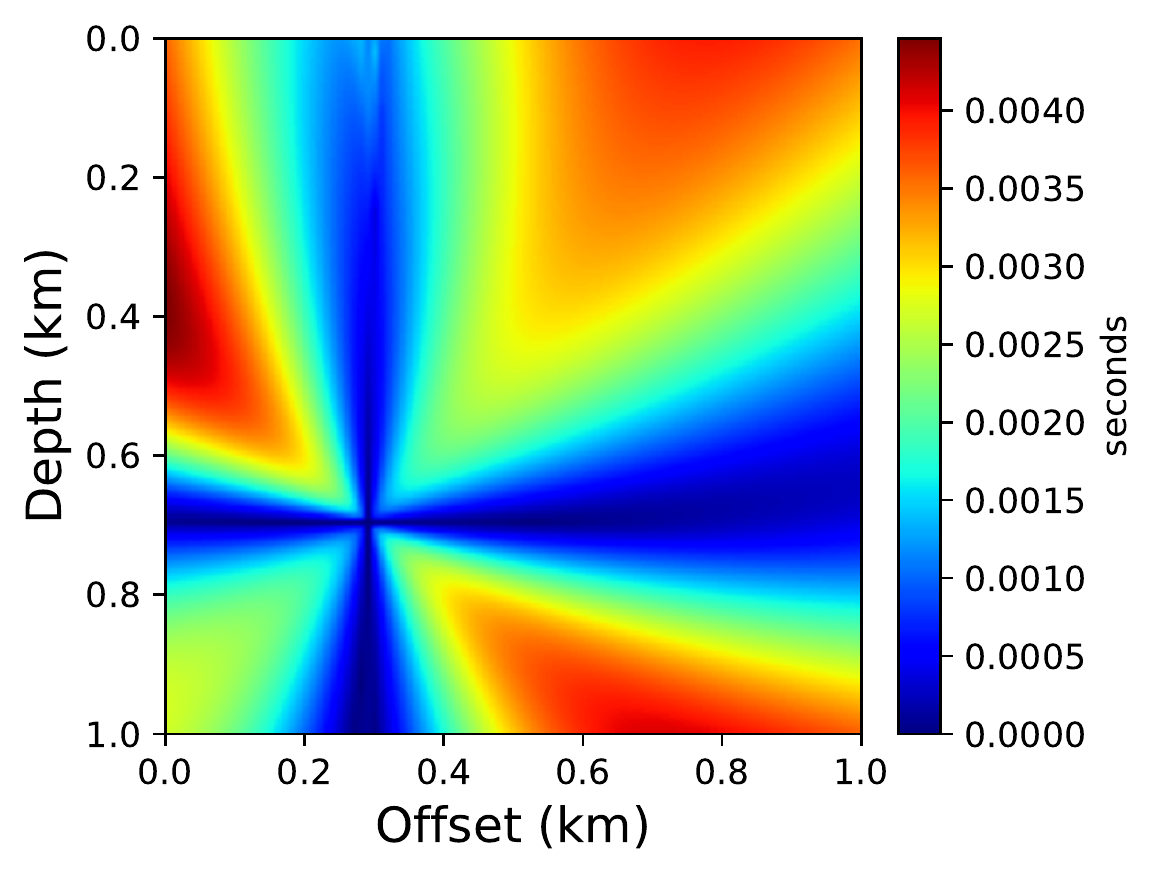}
}
\end{center}
\caption{
The absolute traveltime errors for the solution computed using the surrogate model (a) and the fast sweeping solution (b) for the TTI model considered in example 2 for a randomly chosen source location. 
}%
\label{fig:vofxz_sur_errors}
\end{figure}

\begin{figure}[]
\begin{center}
\includegraphics[width=0.5\textwidth]{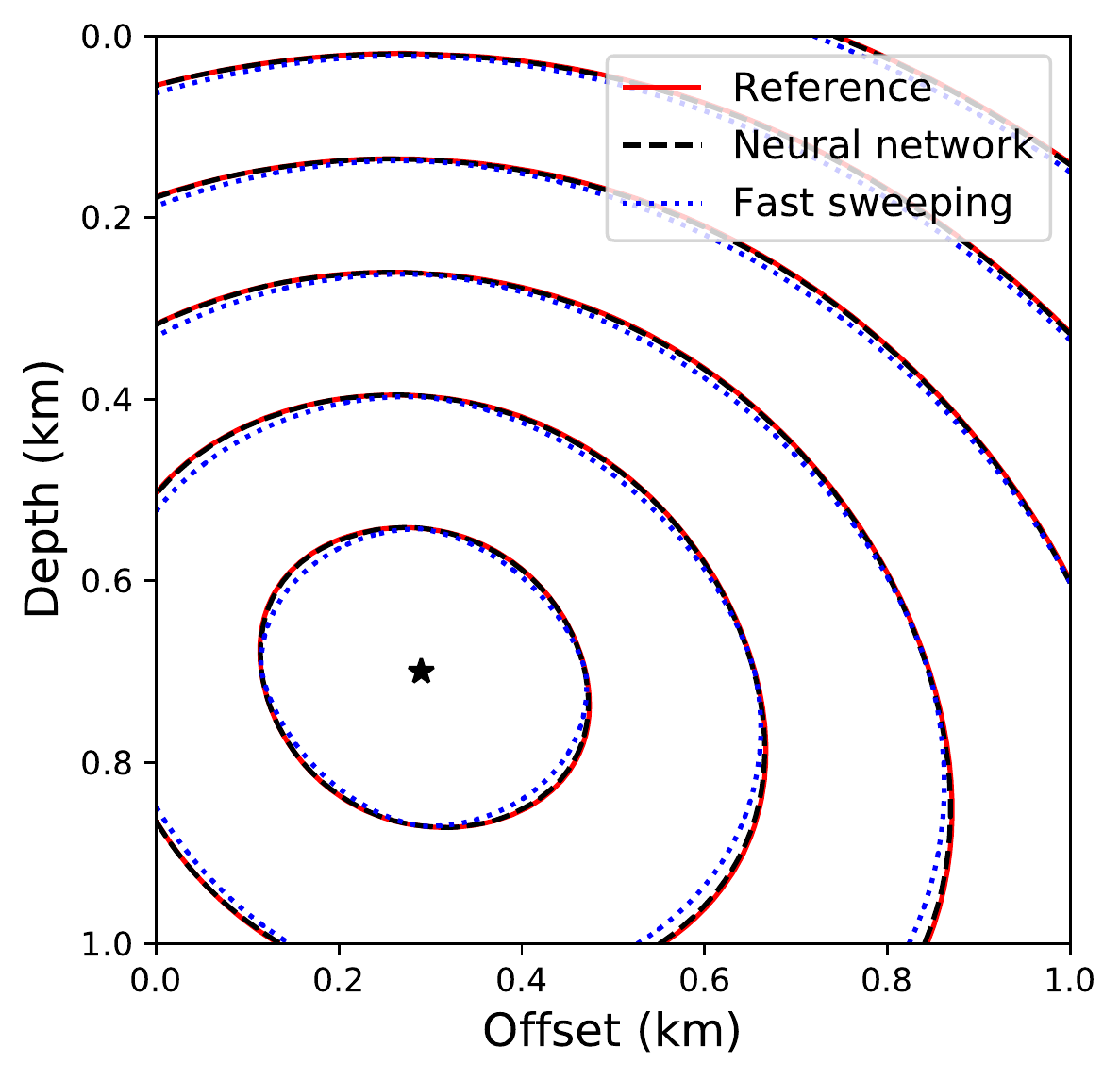}
\end{center}
\caption{
The traveltime contours for solutions obtained using the reference solution (solid red), the neural network surrogate model (dashed black), and the first-order fast sweeping solver (dotted blue) for a randomly chosen source point in a smoothly varying TTI model. 
}%
\label{vofz_surrogate_contours}
\end{figure}

After training the surrogate model, we test its performance by computing the traveltime field corresponding to a randomly chosen source location. Figure~\ref{fig:vofxz_sur_errors} compares the absolute traveltime errors for the solution predicted by the surrogate model and the fast sweeping TTI solver. We observe that even without any additional training for this new source position, we obtain remarkably high accuracy compared to the fast sweeping method. This is also confirmed by visually comparing the corresponding traveltime contours in Figure~\ref{vofz_surrogate_contours}.

\subsection*{Example 3: VTI SEAM model}

\begin{figure}[]
\begin{center}
\subfigure[]{%
\includegraphics[width=0.48\textwidth]{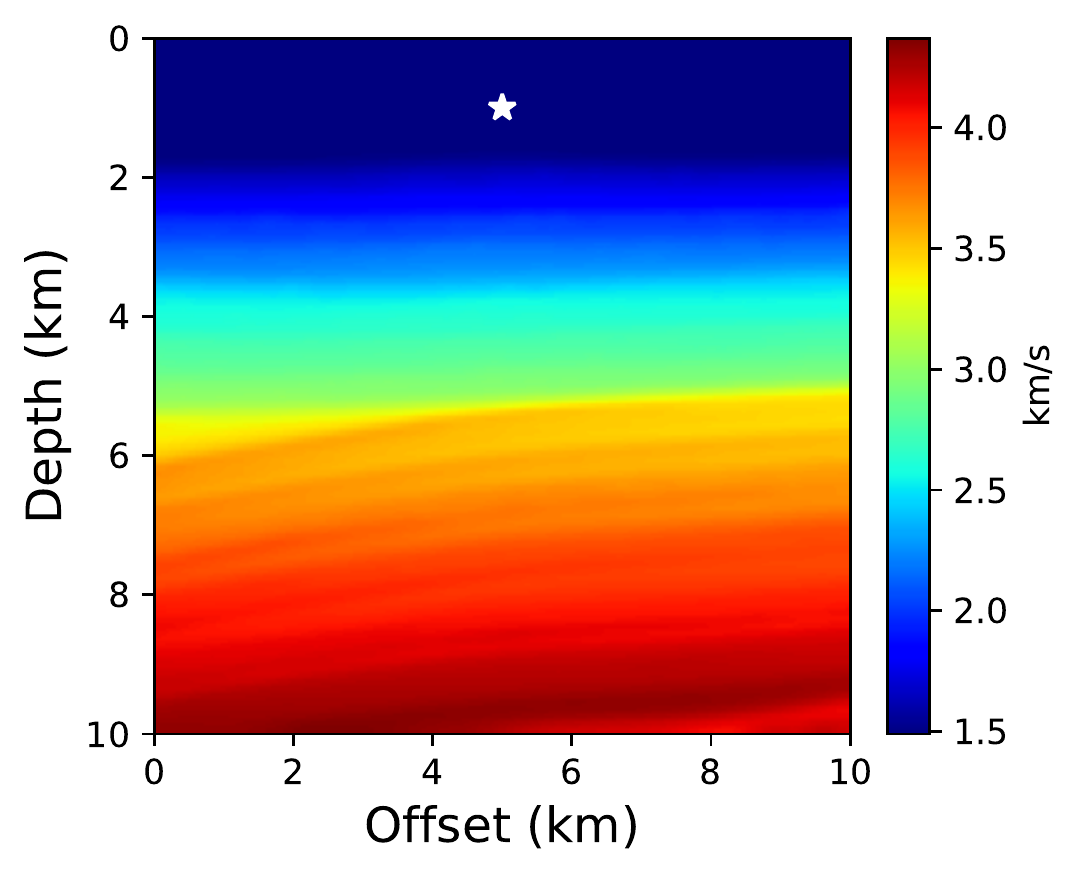}
}
\subfigure[]{%
\includegraphics[width=0.48\textwidth]{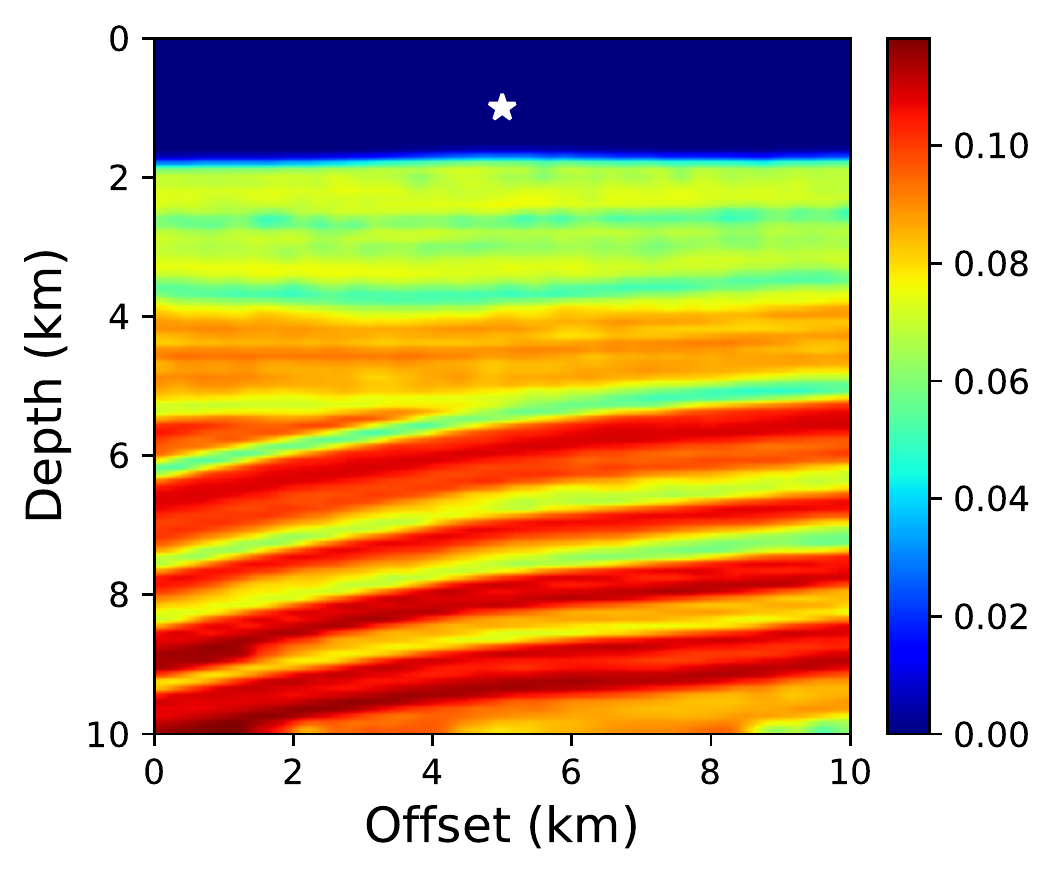}
}
\subfigure[]{%
\includegraphics[width=0.48\textwidth]{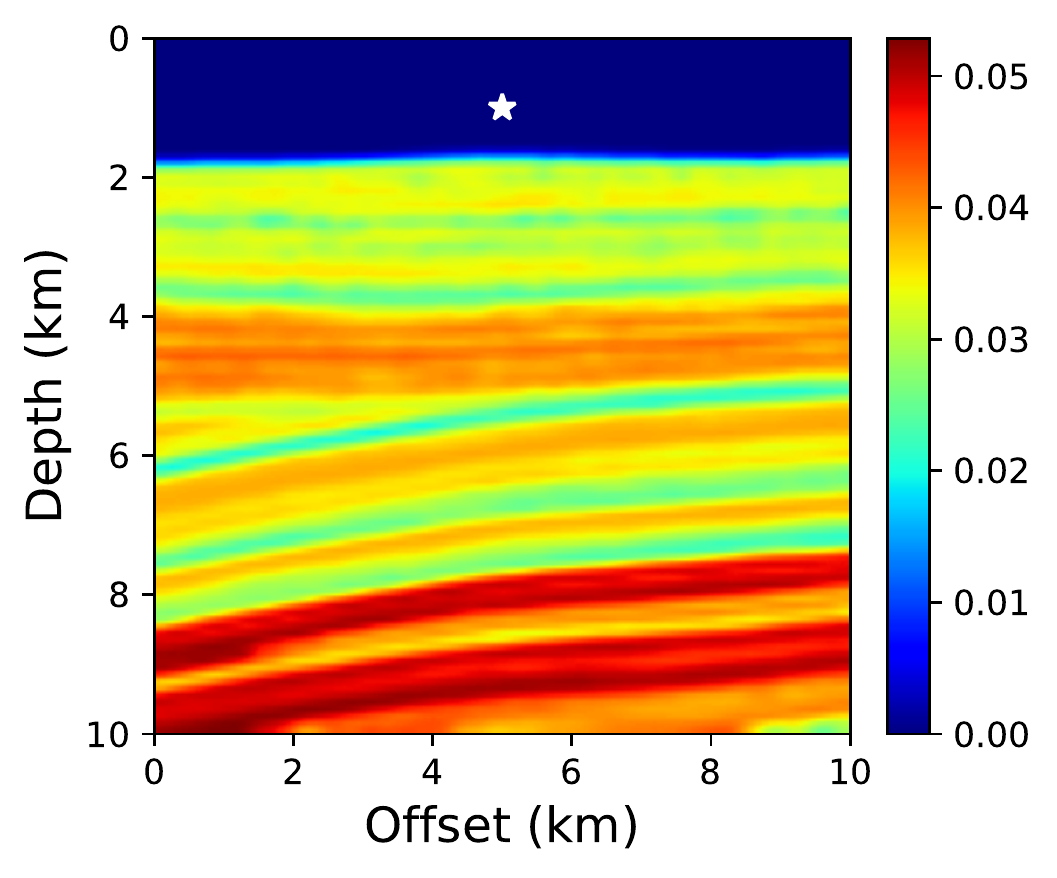}
}
\end{center}
\caption{
(a) The vertical velocity $v_t$, (b) the $\epsilon$ parameter, and (c) the $\eta$ parameter for the considered portion of the VTI SEAM model. The white star indicates the position of the point-source.
}%
\label{fig:vtiseam_model}
\end{figure}

Next, we test the performance of the proposed method on a portion of the VTI SEAM model, shown in Figure~\ref{fig:vtiseam_model}. {\color{black} The model parameters are extracted from the 3D SEG Advanced Modeling (SEAM) Phase I subsalt earth model~\cite[]{fehler2011seam}.} This is a particularly interesting example due to sharper variations in the velocity model and the anisotropy parameters. The model is discretized on a 101$\times$101 grid with a grid spacing of 100 m along both axes. Based on our recent efforts in using the neural network solver~\cite{song2021solving,bin2020eikonal}, we observe that the convergence of the neural network approach slows down considerably in the presence of sharp variations in the velocity model. We have already seen above that using a pre-trained neural network yields faster convergence by allowing the use of the second-order optimization method (L-BFGS-B) directly. Therefore, in this example, we use the pre-trained network obtained from the TTI eikonal solver in example 2.

For further speedup, we propose a two-stage training scheme to obtain an accurate traveltime solution. In the first stage, when the neural network is learning a smooth representation of the underlying function, we use only a small percentage of grid points for training. In this case, we use only 1\% of the total grid points chosen randomly for the first 200 epochs and then switch to using 50\% of the total grid points in stage 2 for another 1000 epochs to update the learned function in better approximating sharp features in the resulting traveltime field. Again, since we start training with a pre-trained network, we use the L-BFGS-B optimizer for faster convergence. 

\begin{figure}[]
\begin{center}
\subfigure[]{%
\label{fig:vtiseam_contours_a}
\includegraphics[width=0.4\textwidth]{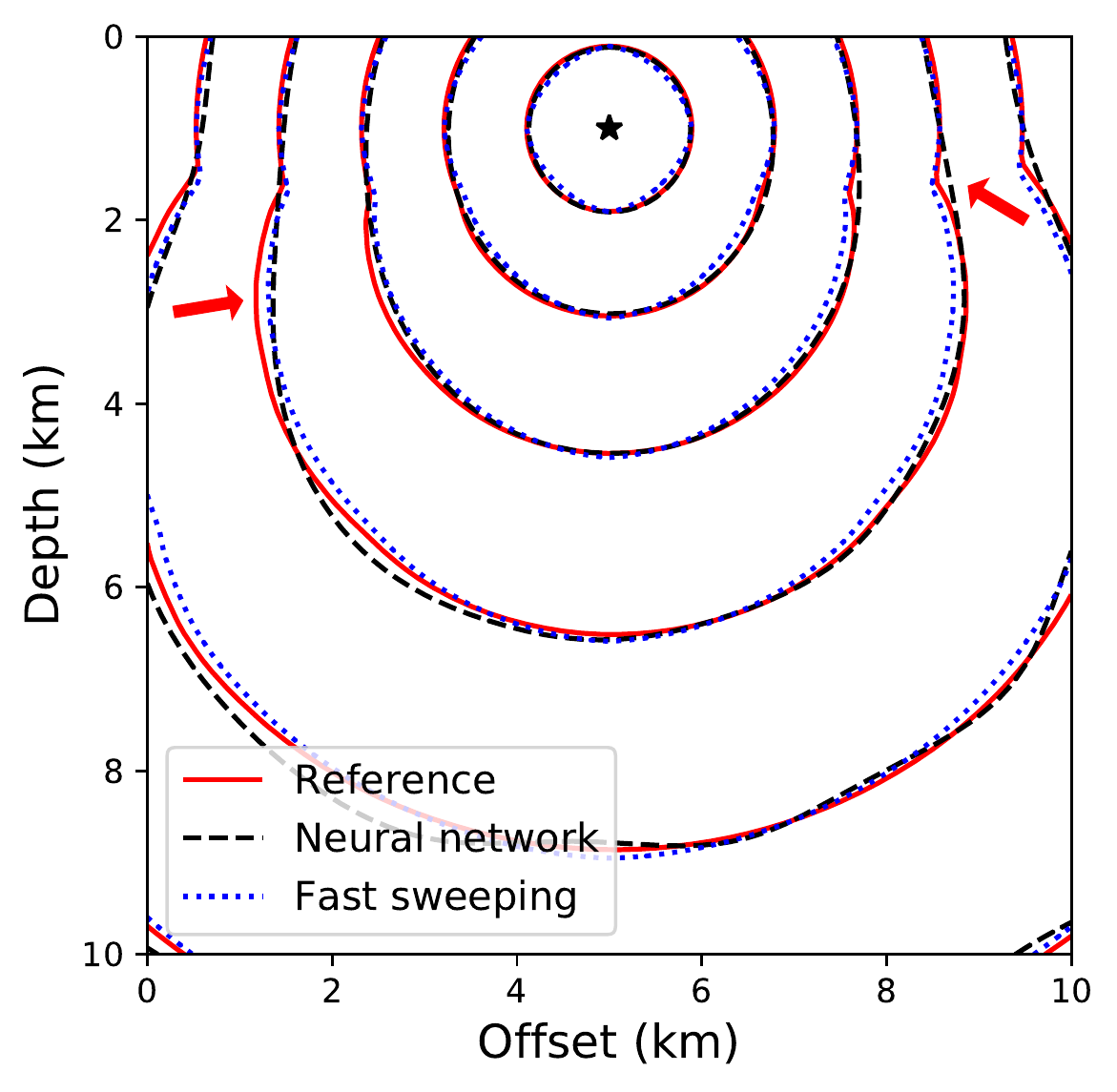}
}
\subfigure[]{%
\label{fig:vtiseam_contours_b}
\includegraphics[width=0.4\textwidth]{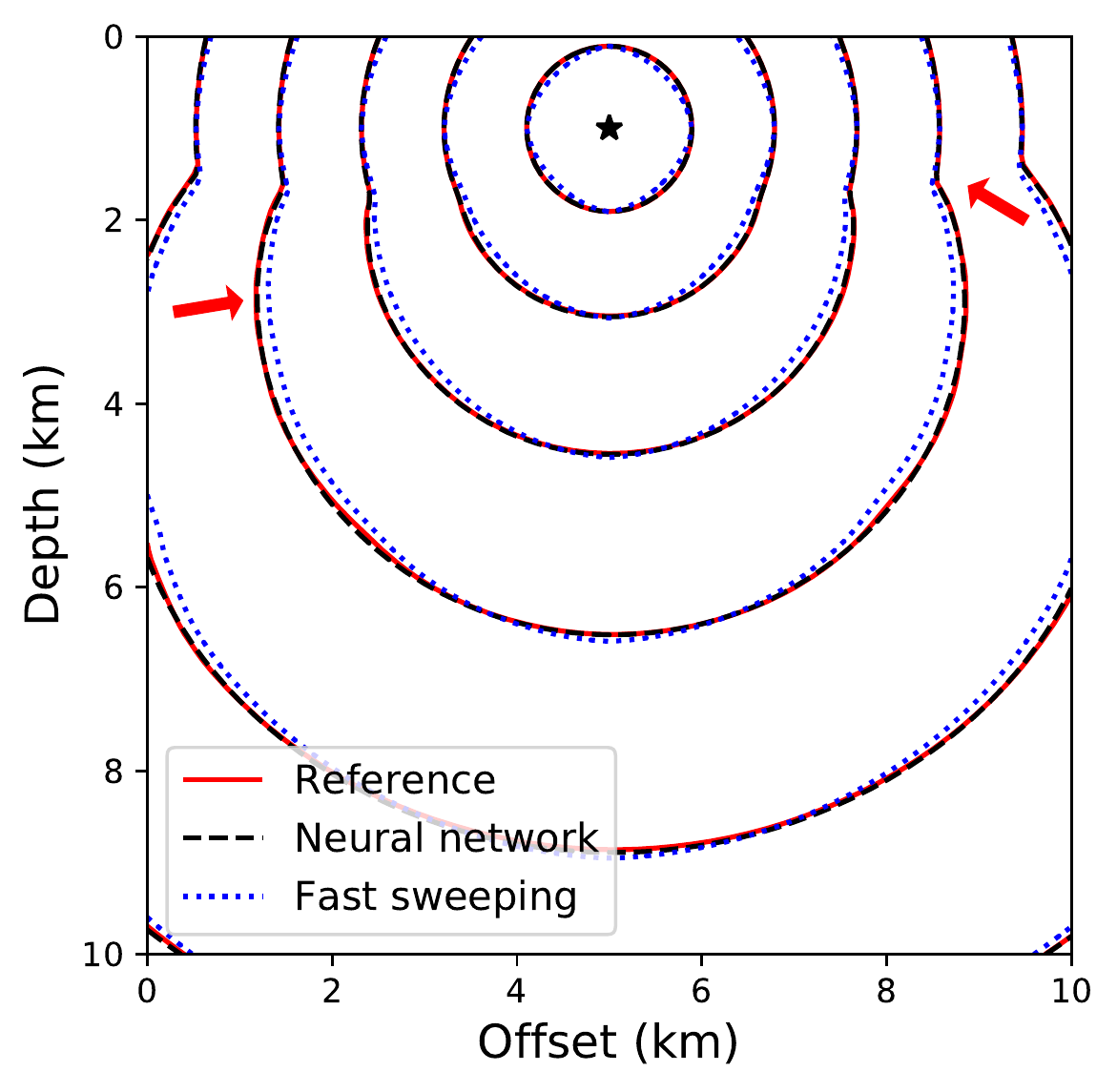}
}
\end{center}
\caption{
The traveltime contours for the reference solution (solid red), neural network solution (dashed black), and the first-order fast sweeping solution (dotted blue) for the VTI SEAM model. Red arrows indicate the improvement of accuracy due to the additional training points used in the neural network solution in going from stage 1 (a) to stage 2 (b) of the proposed training process. The black star shows the location of the point-source. 
}%
\label{fig:vtiseam_contours}
\end{figure}

\begin{figure}[]
\begin{center}
\subfigure[]{%
\includegraphics[width=0.47\textwidth]{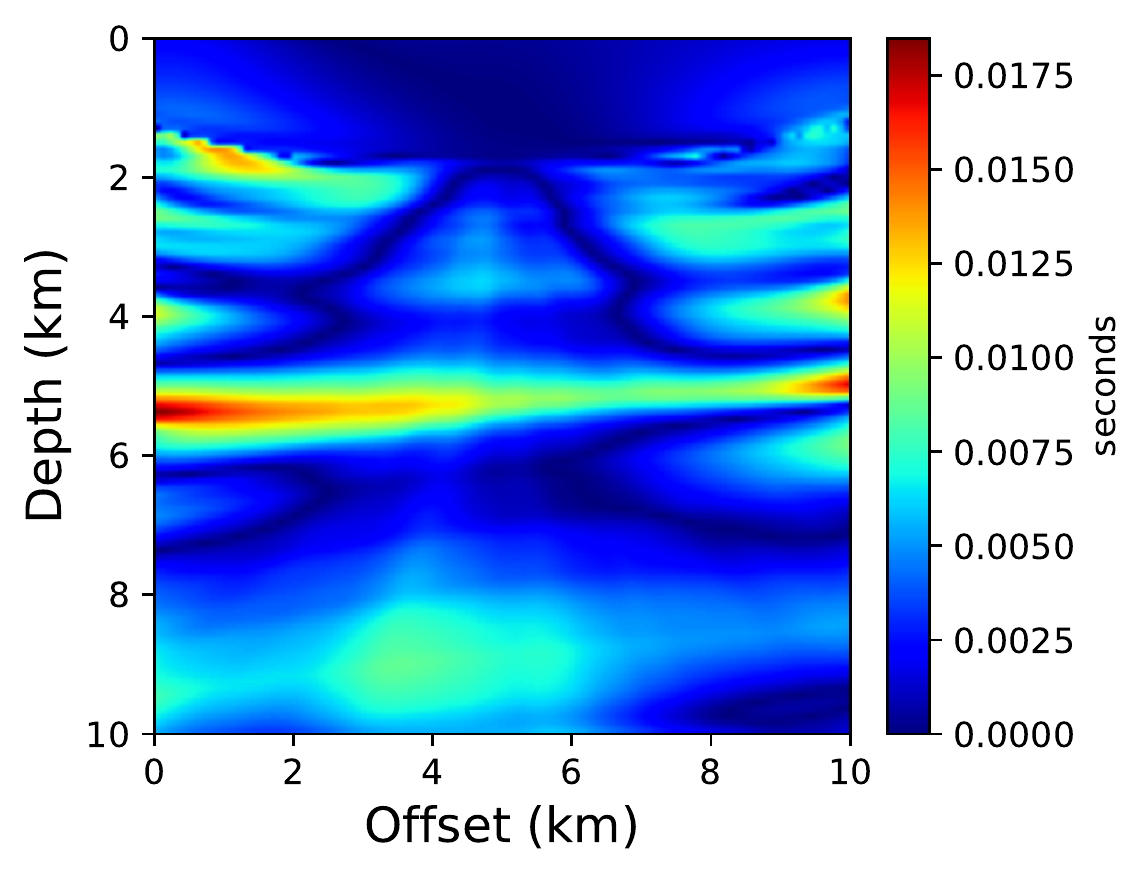}
}
\subfigure[]{%
\includegraphics[width=0.45\textwidth]{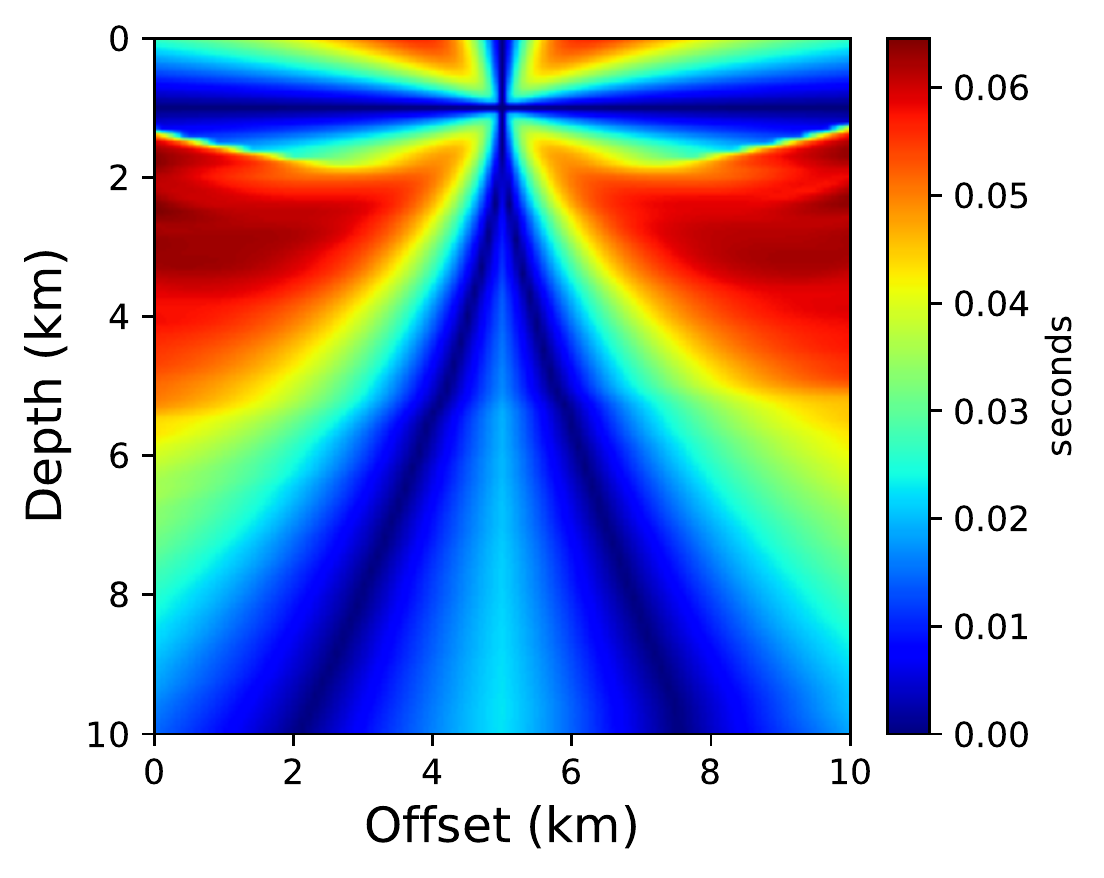}
}
\end{center}
\caption{
The absolute traveltime errors for the neural network solution (a) and the fast sweeping solution (b) for the SEAM VTI model.
}%
\label{fig:vtiseam_errors}
\end{figure}

Figure~\ref{fig:vtiseam_contours} shows traveltime contours for a point-source located at (5~km, 1~km) for the reference solution, the neural network solution, and the first-order fast sweeping solution. In Figure~\ref{fig:vtiseam_contours_a}, we show the neural network solution at the end of stage 1. It can be seen that the solution is quite smooth and misses sharp features visible in the reference solution. In Figure~\ref{fig:vtiseam_contours_b}, we observe that the neural network solution captures these features as additional training points are added in the second stage of training. Therefore, using a small number of training points in stage 1 reduces the training cost without compromising on solution accuracy. By comparing the absolute traveltime errors in Figure~\ref{fig:vtiseam_errors}, we observe that the neural network solution after stage 2 is considerably more accurate than the first-order fast sweeping method, even for a realistic VTI model.

\subsection*{Example 4: BP TTI model}

\begin{figure}[]
\begin{center}
\subfigure[]{%
\includegraphics[width=0.4\textwidth]{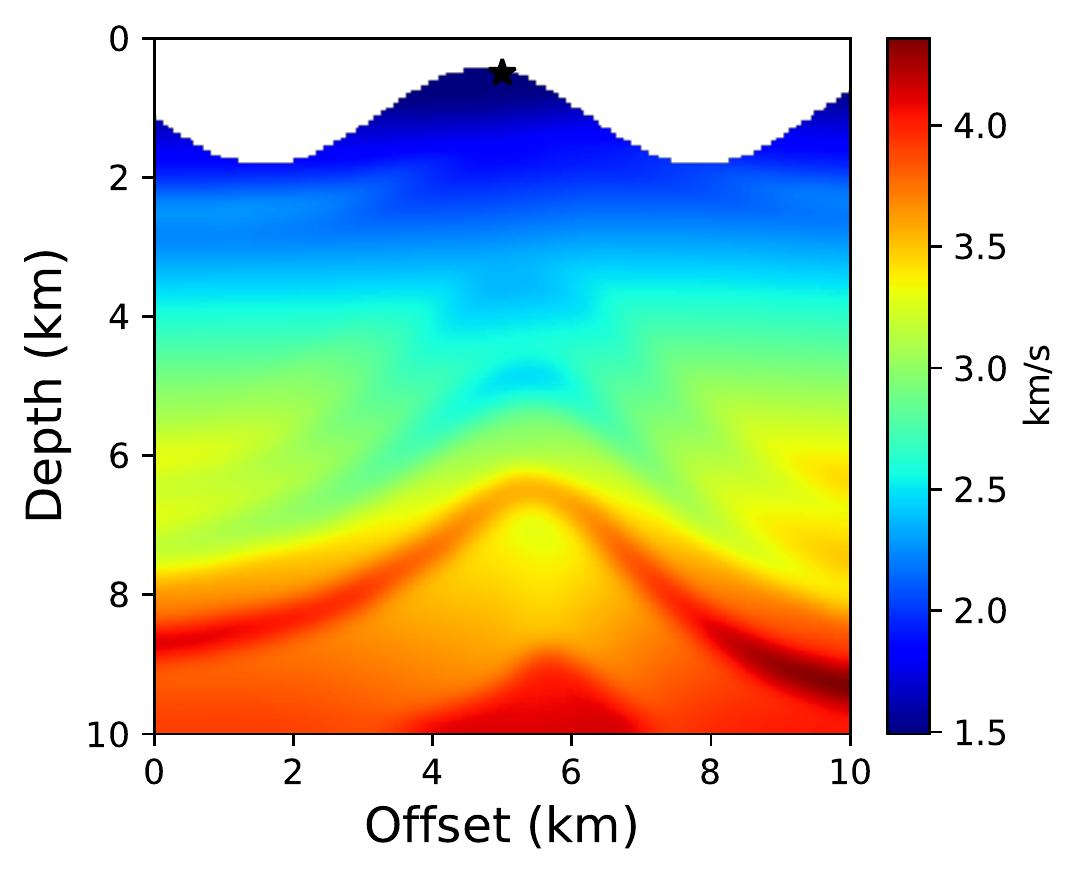}
}
\subfigure[]{%
\includegraphics[width=0.4\textwidth]{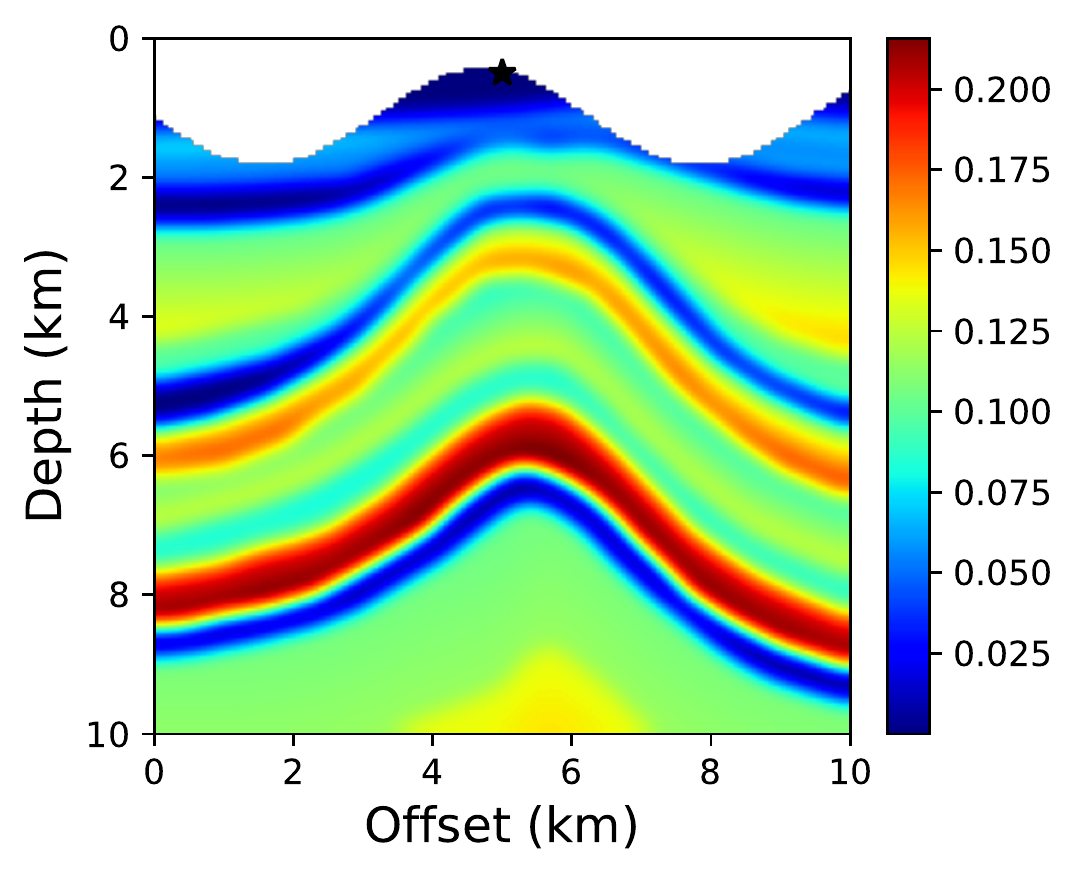}
}
\subfigure[]{%
\includegraphics[width=0.4\textwidth]{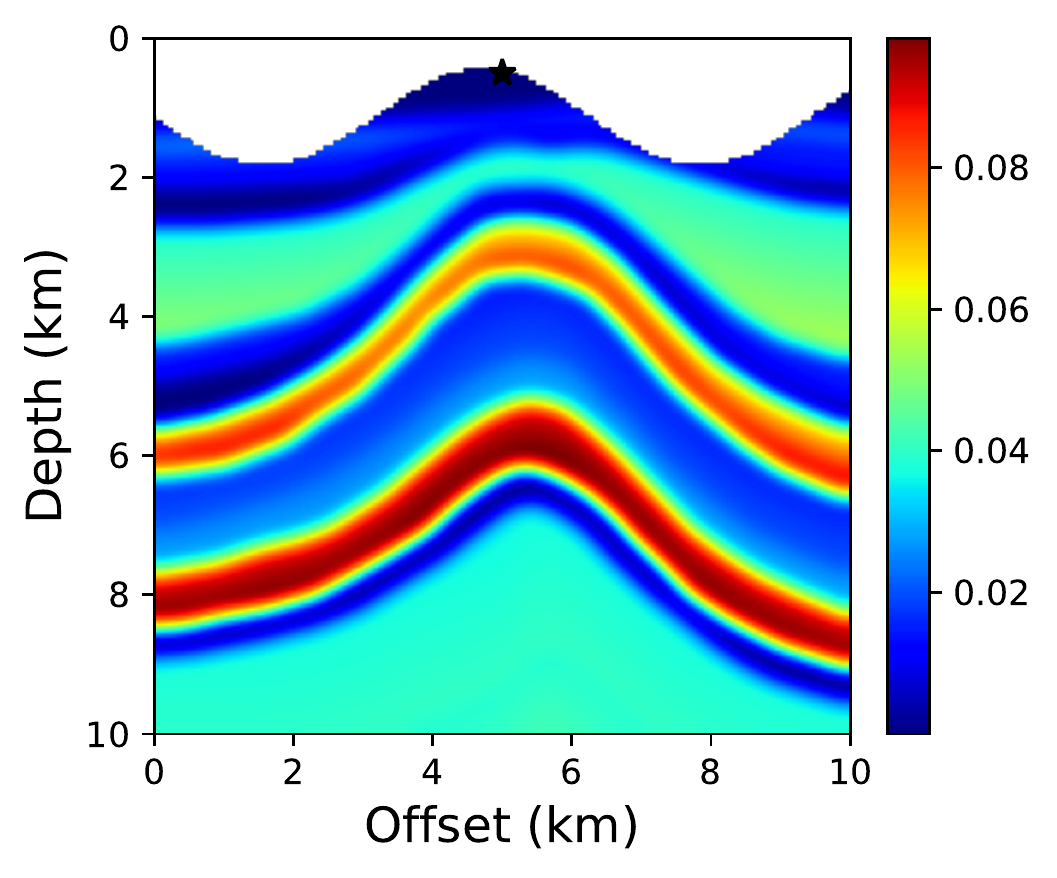}
}
\subfigure[]{%
\includegraphics[width=0.4\textwidth]{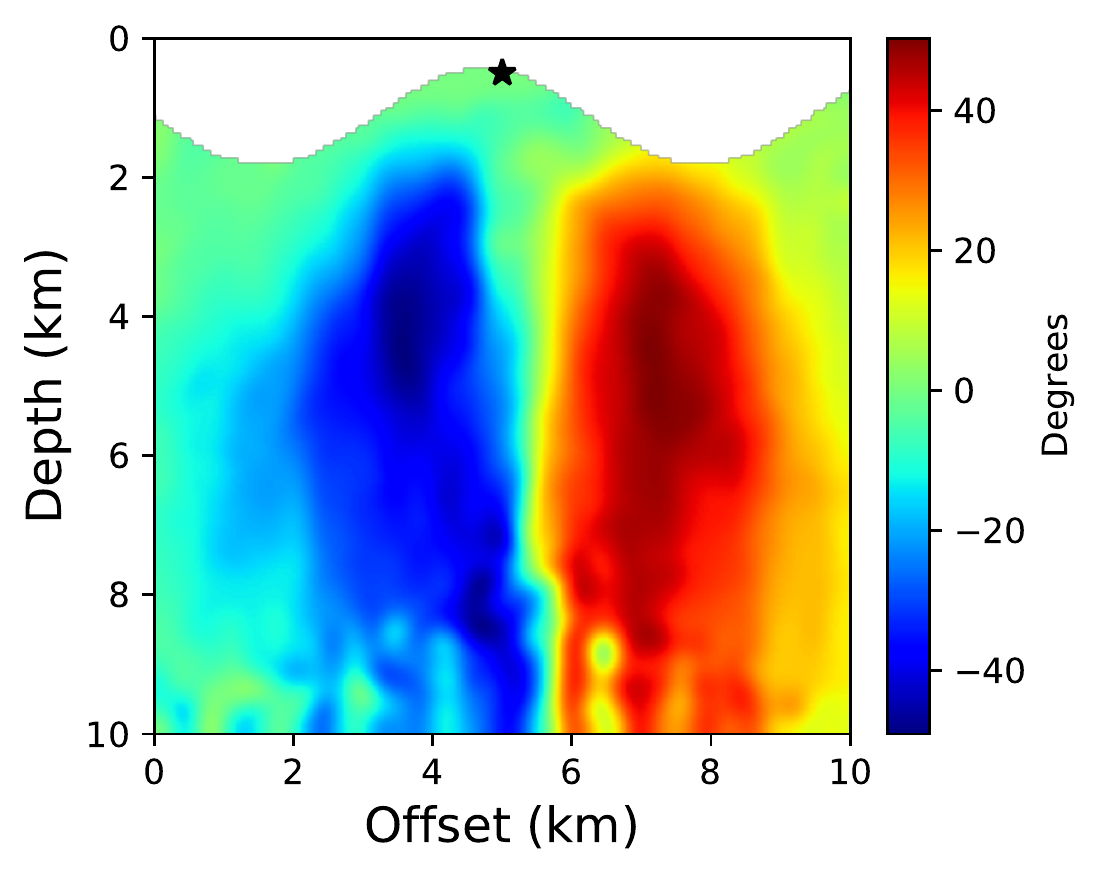}
}
\end{center}
\caption{
(a) The velocity along the symmetry axis $v_t$, (b) the $\epsilon$ parameter, (c) the $\eta$ parameter, and (d) the tilt angle $\theta$ for the considered portion of the BP TTI model with a layer of non-flat topography. 
}%
\label{fig:bptti_model}
\end{figure}

\begin{figure}[]
\begin{center}
\subfigure[]{%
\includegraphics[width=0.47\textwidth]{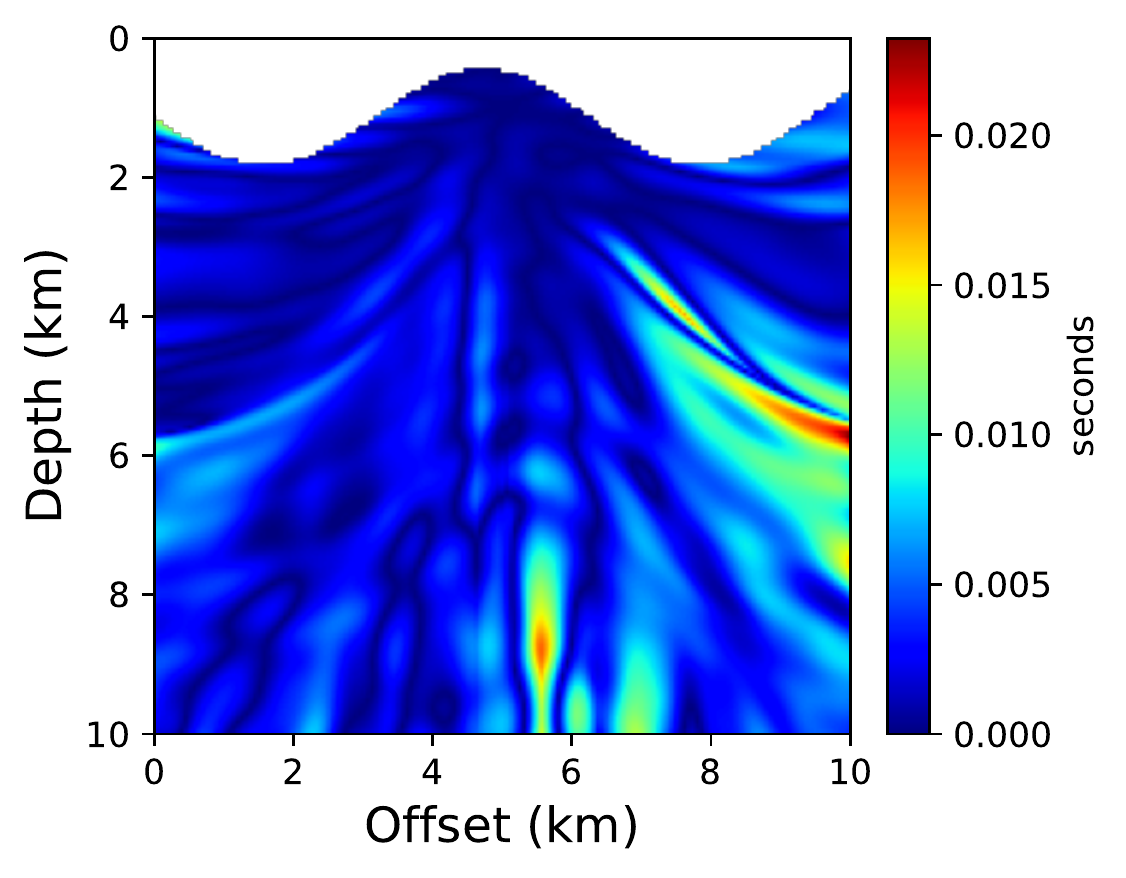}
}
\subfigure[]{%
\includegraphics[width=0.445\textwidth]{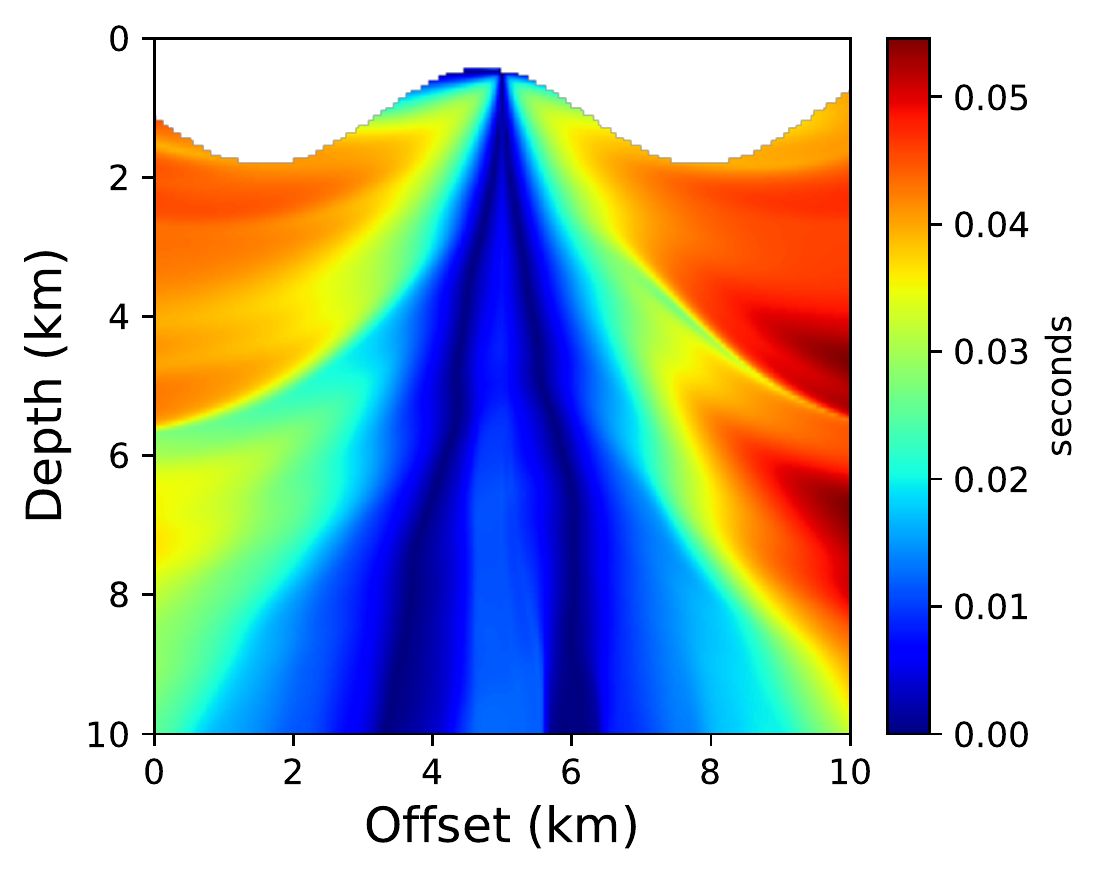}
}
\end{center}
\caption{
The absolute traveltime errors for the neural network solution (a) and the fast sweeping solution (b) for the BP TTI model.
}%
\label{fig:bptti_errors}
\end{figure}

We have already seen how the neural network based approach is flexible in incorporating complex physics compared to conventional techniques. In this example, we will see how incorporating irregular topography is straightforward using the proposed approach. The free-surface encountered in land seismic surveys is often non-flat and requires taking this into account for accurate traveltime computation. One approach to tackle this is to transform from Cartesian to curvilinear coordinate system to mathematically flatten the free-surface and solve the resulting topography-dependent eikonal equation~\cite{lan2013high}. This adds additional computational cost and may result in instabilities when topography varies sharply. On the contrary, the neural network approach outlined here is mesh-free and doesn't require any modification to the algorithm. We demonstrate this through a test on a portion of the BP TTI model, shown in Figure~\ref{fig:bptti_model}. {\color{black} The model was developed by Shah~\cite[]{shah20072007} and publicly provided courtesy of the BP Exploration Operating Company.} The considered portion of the model is discretized on a 161$\times$161 grid using a grid spacing of 6.25~m along both axes. Points above the considered topography layer are then removed from the model.

\begin{figure}[]
\begin{center}
\includegraphics[width=0.5\textwidth]{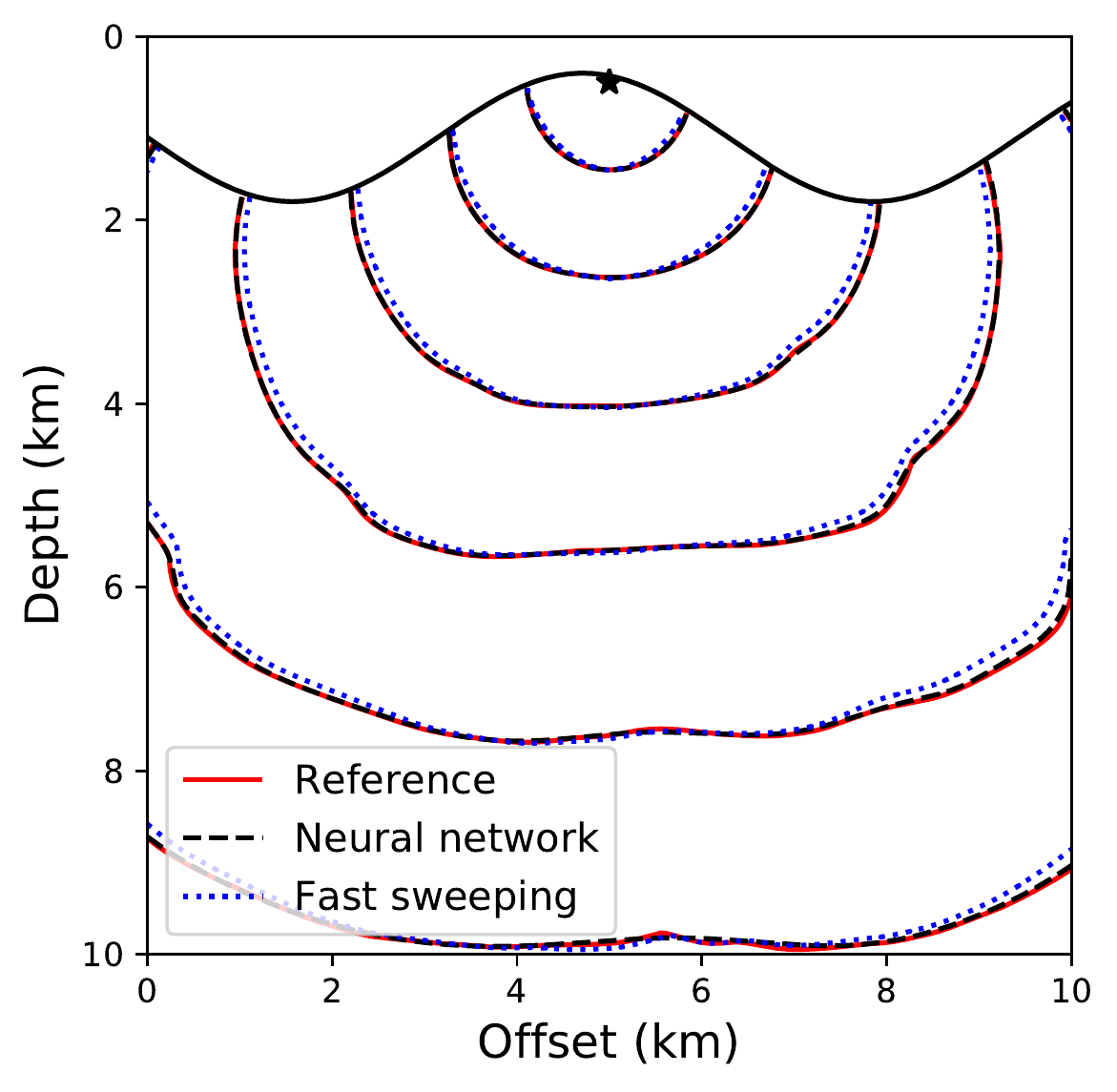}
\end{center}
\caption{
The traveltime contours for the reference solution (solid red), neural network solution (dashed black), and the first-order fast sweeping solution (dotted blue) for the BP TTI model. The black star shows the location of the point-source and the solid black curve indicates the topography layer.
}%
\label{fig:bptti_contours}
\end{figure}

For a point-source located at (5~km, 0.5~km), we compare the neural network and fast sweeping traveltime solutions. For training the neural network, we take into account about 50\% of the grid points below the topography and once the network is trained, we evaluate the solution on the regular grid points that fall below the topography. We start training using pre-trained parameters from example 2 and train for 10,000 epochs using an L-BFGS-B optimizer. Figure~\ref{fig:bptti_errors} shows absolute traveltime errors for the two cases, indicating that the neural network solution is again considerably more accurate than fast sweeping. We confirm this observation visually through traveltime contours plotted in Figure~\ref{fig:bptti_contours}.

\subsection*{Example 5: 3D TTI model}

\begin{figure}[]
\begin{center}
\includegraphics[width=0.7\textwidth]{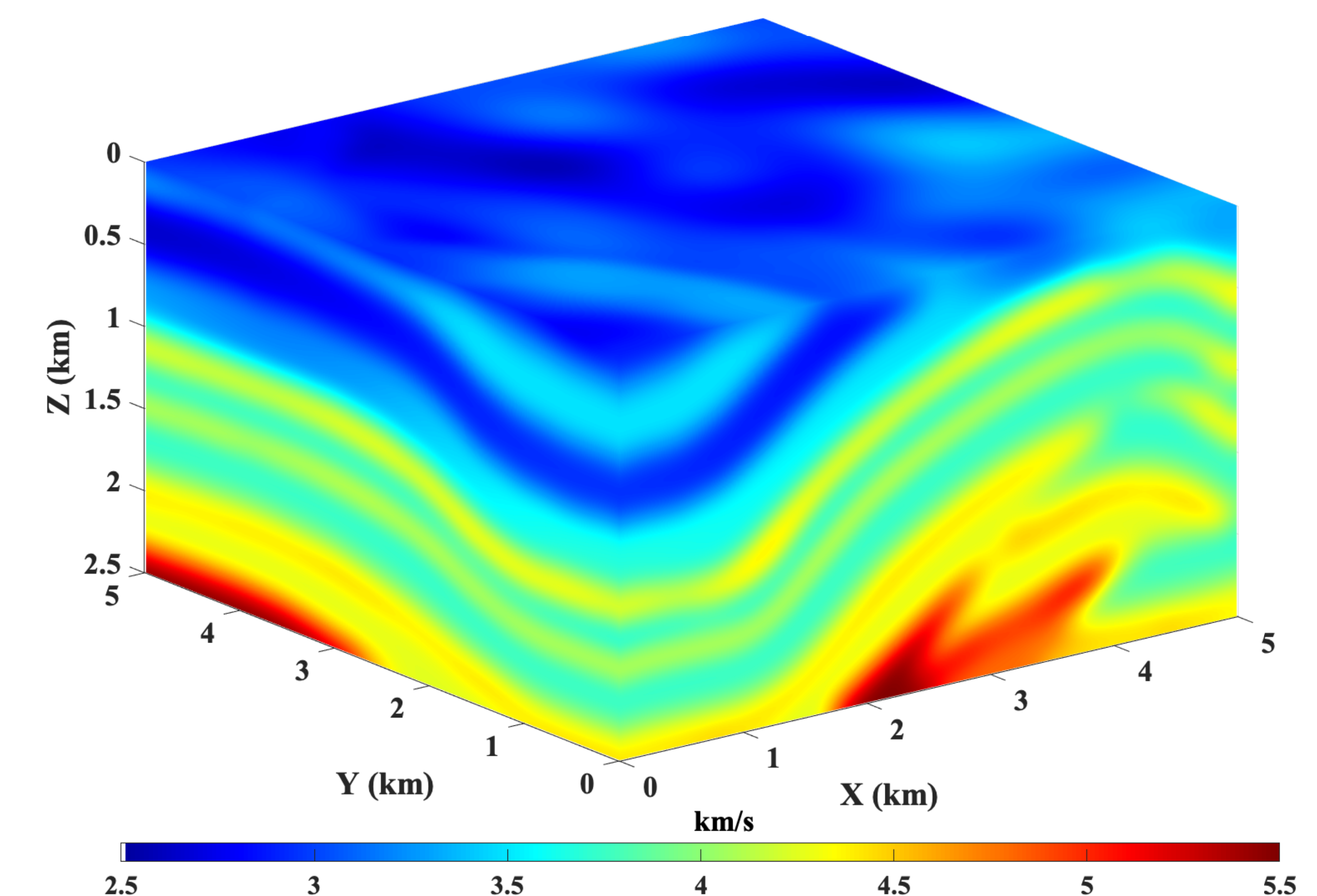}
\end{center}
\caption{
A velocity model for the parameter $v_t$ for the 3D TTI model test. A homogeneous model is used for the anisotropy parameters ($\epsilon$=0.2, $\eta$ = 0.1) and for the tilt angle ($\theta$ = 30$^\circ$). 
}%
\label{fig:overthrust_vz}
\end{figure}

Finally, we show an example of extending the proposed method to a 3D TTI model. The model for the velocity parameter $v_t$ is shown in Figure~\ref{fig:overthrust_vz}. A homogeneous model is used for the anisotropy parameters with $\epsilon$=0.2 and $\eta$ = 0.1. We also consider a homogeneous tilt angle of 30$^\circ$. The model is discretized on a 101$\times$101$\times$51 grid with a grid spacing of 50~m along each axis. The workflow for obtaining the neural network solution is essentially the same. In this case, the neural network takes three input parameters corresponding to the spatial axes $(x,y,z)$ and outputs the corresponding unknown traveltime field $\hat{\tau}(x,y,z)$. Similar to before, the neural network output is multiplied by the known traveltime factor $T_0(x,y,z)$ to obtain the final traveltime solution. 

\begin{figure}[]
\begin{center}
\subfigure[]{%
\includegraphics[width=0.48\textwidth]{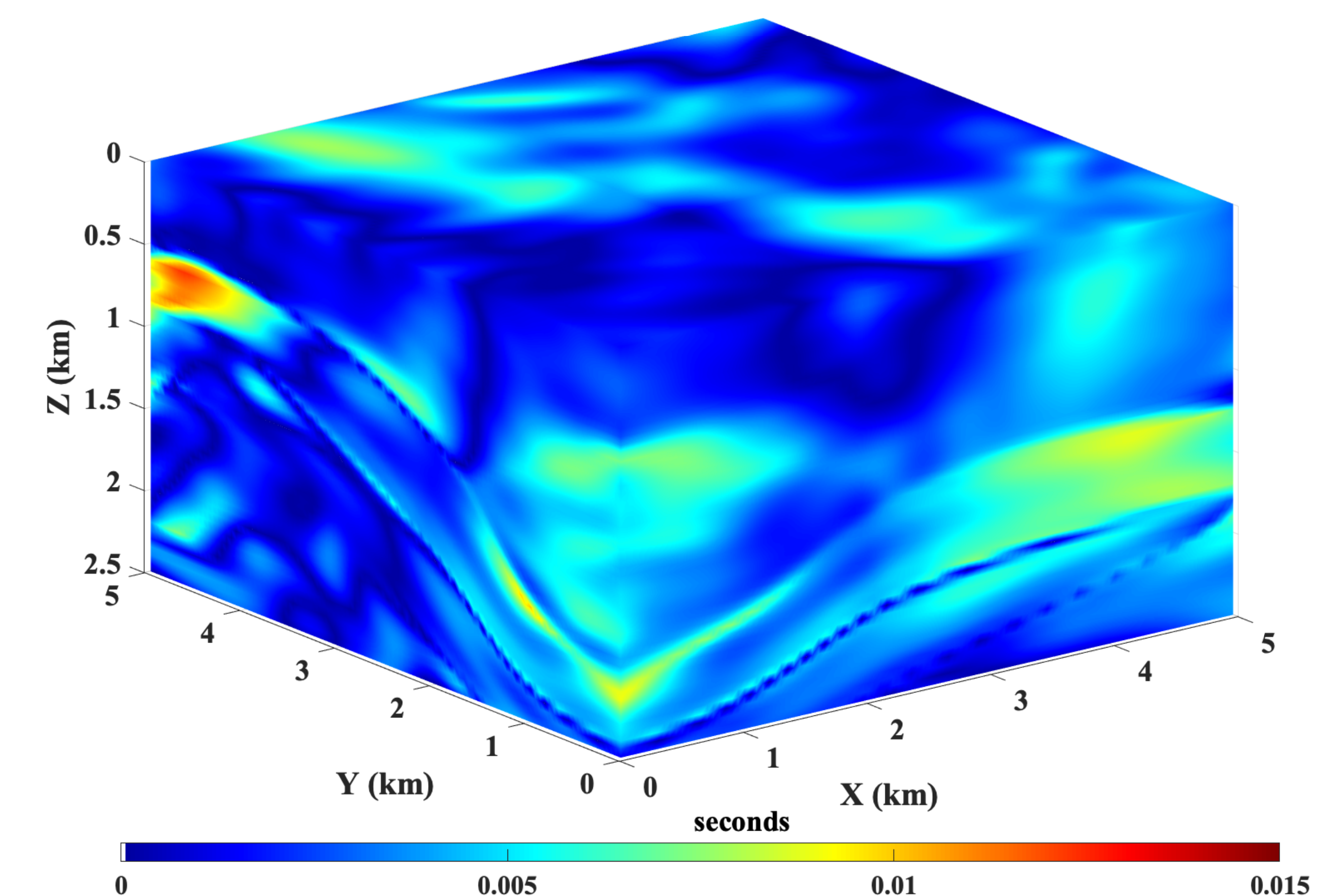}
}
\subfigure[]{%
\includegraphics[width=0.48\textwidth]{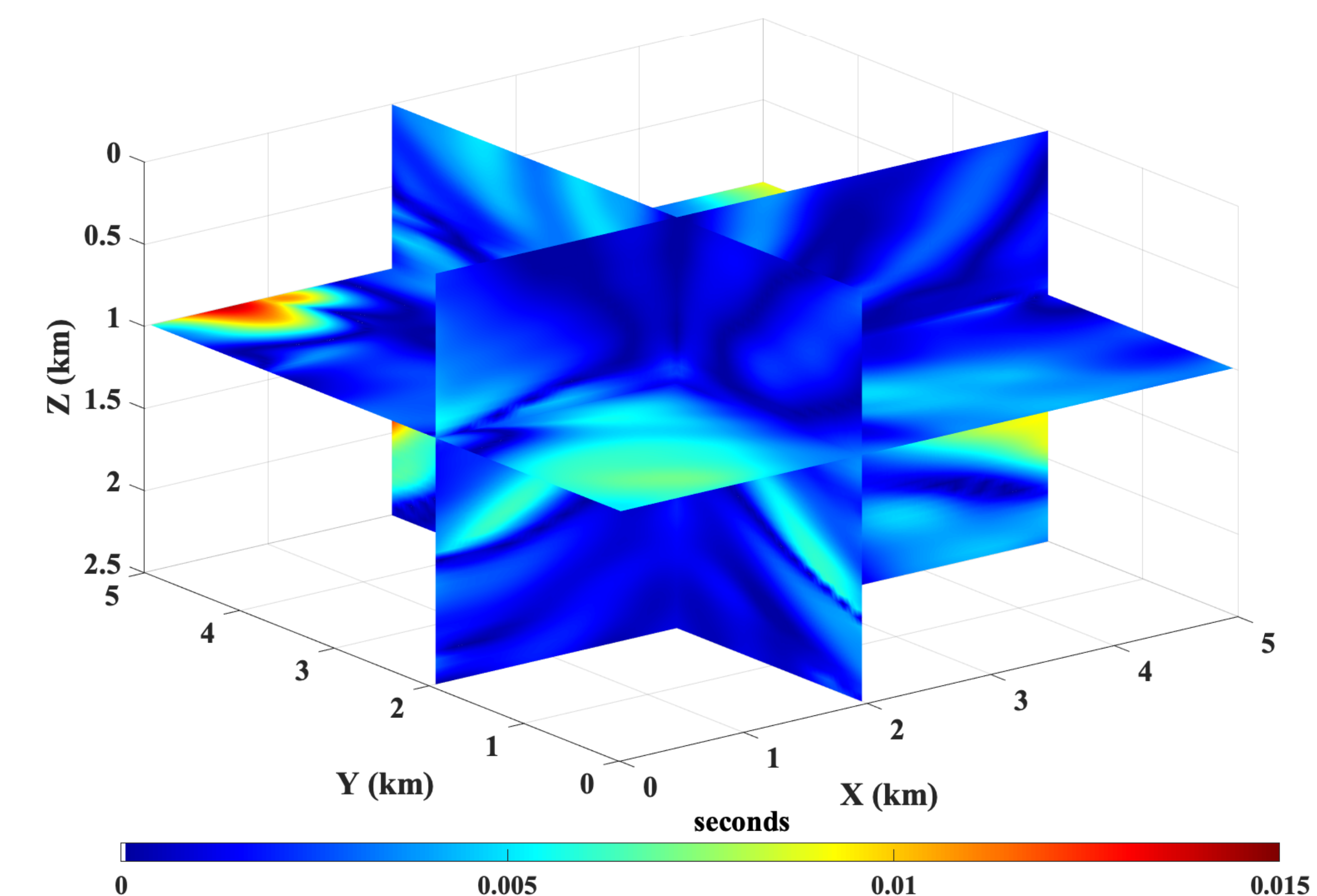}
}
\subfigure[]{%
\includegraphics[width=0.48\textwidth]{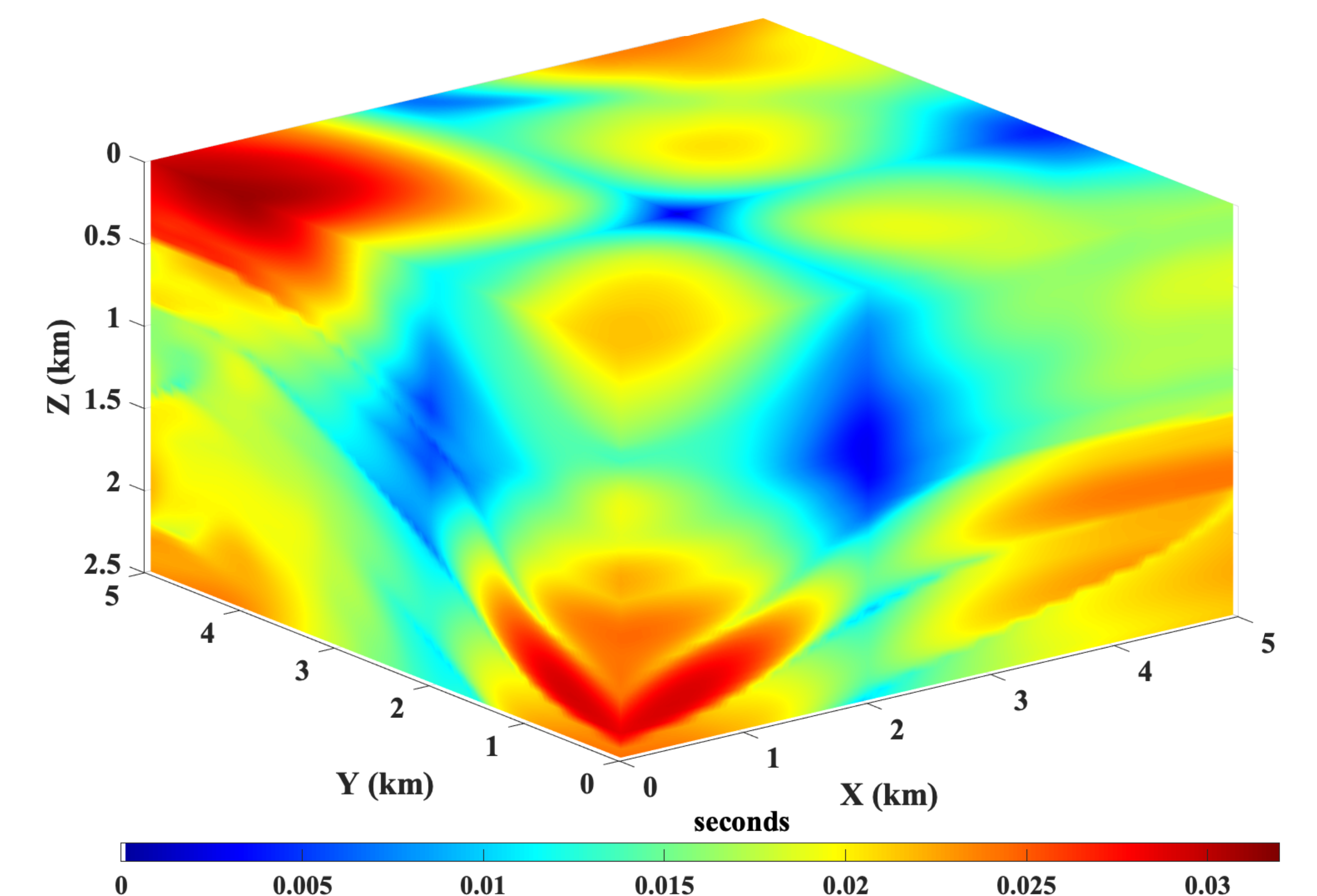}
}
\subfigure[]{%
\includegraphics[width=0.48\textwidth]{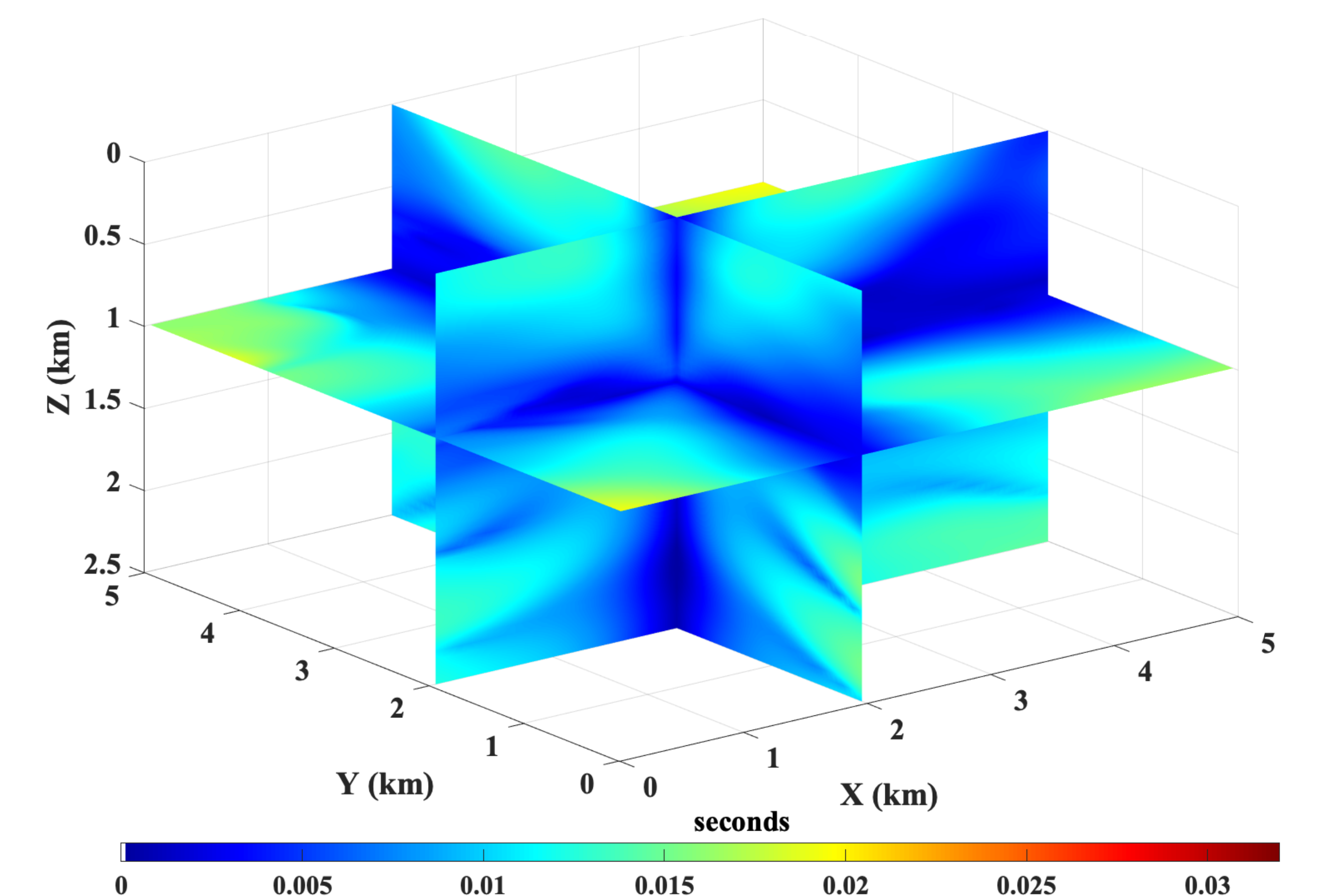}
}
\end{center}
\caption{
The absolute traveltime errors for the neural network solution (a,b) and the fast sweeping solution (c,d) for the 3D TTI model.
}%
\label{fig:overthrust_errors}
\end{figure}

Starting with a pre-trained neural network on a smoothly varying TTI model, we train the network using 50\% of the total grid points chosen randomly. We use 20,000 L-BFGS-B epochs during the training process. For a point-source located at $(x,y,z)$=(2~km, 2~km, 1~km), we compare the accuracy of the neural network solution and the first-order fast sweeping solution in Figure~\ref{fig:overthrust_errors}. We observe that the proposed method is capable of computing accurate traveltimes for 3D TTI models as well without requiring any major alteration to the underlying algorithm.

\section{Discussion and conclusions}

We proposed a neural network approach to computing first-arrival traveltimes based on the framework of physics-informed neural networks. By leveraging the capabilities of neural networks as universal function approximators, we define a loss function to minimize the residual of the governing eikonal equation at a chosen set of training points. This is achieved with a simple feed-forward neural network using automatic differentiation. 

We demonstrated the flexibility of the proposed framework in incorporating anisotropy in the model simply by updating the loss function of the neural network according to the underlying PDE. Since the method is mesh-free, we also saw how easy it is to incorporate non-flat topography into the solver compared to conventional methods. Another attractive feature, due to this mesh-free nature of the algorithm, is that sources and receivers do not have to lie on a regular grid as required by conventional finite-difference techniques. We also showed that by using machine learning techniques like transfer learning and surrogate modeling, we could transfer information gained by solving one problem to the next -- a key feature missing in our conventional numerical algorithms. This would be key in achieving computational efficiency beyond conventional methods on models of practical interest. 

Furthermore, the neural network based eikonal solver uses \texttt{Tensorflow} at the backend, which allows for easy deployment of computations across a variety of platforms (CPUs, GPUs) and architectures (desktops, clusters). On the contrary, significant effort is needed in adapting conventional algorithms to benefit from different computational platforms or architectures. 

In short, the approach tackles many problems associated with obtaining fast and accurate traveltimes, that required decades of research, in a holistic manner. In fact, it opens new possibilities in solving complex forms of the eikonal equation that have remained unsolved by conventional algorithms or solved through some approximation techniques at best. {\color{black}Our recent tests have demonstrated success in solving such equations using neural networks without approximations. This includes solutions to the qSV eikonal equation for anisotropic media~\cite[]{waheed2021qsv} and the attenuating VTI eikonal equation~\cite[]{taufik2021}.}


{\color{black} It is also worth emphasizing that the actual computational advantage of the proposed method, compared to conventional numerical solvers, depends on many factors including the network architecture, optimization hyper-parameters, and sampling techniques. If the initialization of the network and the optimizer learning rate are chosen carefully, the training can be completed quite efficiently. Furthermore, the activation function used, the adaptive weighting of the loss function terms, and the availability of second-order optimization techniques can accelerate the training significantly. Therefore, a detailed study is needed to quantify the computational gains afforded by the proposed neural network-based solver compared to conventional algorithms by considering the afore-mentioned factors. A rudimentary analysis performed in~\cite[]{bin2020eikonal} indicates that once the surrogate model is obtained, the PINN eikonal solver is computationally faster by more than an order of magnitude than the fast sweeping method. Since the computational complexity of solving the anisotropic eikonal equation using conventional methods increases dramatically compared to the isotropic case~\cite{waheed2015iterative}, the proposed framework is computationally more attractive for traveltime modeling in anisotropic media.}

Nevertheless, there are a few challenges associated with the method that requires further research. Chief among them is the slow convergence of the solution in the presence of sharp heterogeneity in the velocity and/or anisotropy models. We proposed a two-stage optimization process in this chapter that alleviates part of the problem by using only a small fraction of the training points during the initial training phase. Since at this stage, the network is learning a smooth representation of the underlying function, we could save some computational cost by using a small number of training points initially. We also used a locally adaptive activation function that has been shown to achieve faster convergence. Other possible solutions may include an adaptive sampling of the velocity model by using denser sampling of training points around parts of the model with large velocity gradients. Another challenge concerns the optimal choice of the neural networks' hyper-parameters. In this study, we alleviated the problem by choosing them through some quick initial tests and keeping them fixed for all the test examples. Recent advances in the field of meta-learning may, potentially, enable automated selection of these parameters in the future.

\newpage





\bibliographystyle{elsarticle-num-names}
\bibliography{pinn_ani_eikonal.bib}

\end{document}